\documentclass[showpacs,amsmath,amssymb,aps,prd, preprint,groupedaddress,superscriptaddress,nofootinbib]{revtex4-2} 
\usepackage[utf8]{inputenc}
\usepackage{graphicx}
\usepackage{tensor}
\usepackage{bm}
\usepackage{color}

\newcommand{\rcd}[1]{\tensor{\mathring{\nabla}\!}{#1}}
\newcommand{\leviconnection}[1]{\tensor{\mathring{\bm{\nabla}}\!}{#1}}
\newcommand{\rsconnection}[1]{\tensor{\mathring{\omega}}{#1}}

\newcommand{\potential}[1]{\tensor{\Sigma}{#1}}
\newcommand{\pd}[1]{\tensor{\partial}{#1}}
\newcommand{\nablab}{\bm{\nabla}}

\newcommand{\e}[2][]{\tensor{#1{e}}{#2}}

\newcommand{\teta}[2][]{\tensor{#1\vartheta}{#2}}

\newcommand{\lorentz}[2][]{\tensor{#1\Lambda}{#2}}

\newcommand{\bracket}[1]{\left<#1\right>}
\newcommand{\torsion}[1]{\tensor{T}{#1}}

\newcommand{\energy}[1]{\tensor{t}{#1}}

\newcommand{\sech}{\textrm{sech}}
\newcommand{\acceleration}[2][]{\tensor{#1\phi}{#2}}

\DeclareMathOperator{\arcsinh}{arcsinh}
\DeclareMathOperator{\arctanh}{arctanh}

\begin{document}


\title{On the gravitational energy problem and the energy of photons}



\author{J. B. Formiga}
\email[]{jansen@fisica.ufpb.br}
\affiliation{Departamento de Física, Universidade Federal da Paraíba, Caixa Postal 5008, 58051-970 João Pessoa, Pb, Brazil}

\author{Duarte, Jo\~ao A. C.}

\date{\today}


\date{\today}

\begin{abstract}
The lack of a well-established solution for the gravitational energy problem might be one of the reasons why a clear road to quantum gravity does not exist. In this paper, the gravitational energy is studied in detail with the help of the teleparallel approach that is equivalent to general relativity. This approach is applied to the solutions of the Einstein-Maxwell equations known as $pp$-wave spacetimes. The quantization of the electromagnetic energy is assumed and it is shown that the proper area measured by an observer must satisfy an equation for consistency. The meaning of this equation is discussed and it is argued that the spacetime geometry should become discrete once all matter fields are quantized, including the constituents of the frame; it is shown that for a harmonic oscillation with wavelength $\lambda_0$, the area and the volume take the form $A=4(N+1/2)l_p^2/n$ and $V=2(N+1/2)l_p^2\lambda_0$, where $N$ is the number of photons, $l_p$ the Planck length, and $n$ is a natural number associated with the length along the $z$-axis of a box with cross-sectional area $A$. The localization of the gravitational energy problem is also discussed. The stress-energy tensors for the gravitational and electromagnetic fields are decomposed into energy density, pressures and heat flow. The resultant expressions are consistent with the properties of the fields, thus indicating that one can have a well-defined energy density for the gravitational field regardless of the principle of equivalence.
\end{abstract}


\maketitle

\section{Introduction} 
There exist two problems in gravity that have not been solved yet and are frequently treated as uncorrelated, namely, a full description of the gravitational energy and a full quantum theory of gravity. Although apparently uncorrelated, they both have something in common, their tension with the principle of equivalence \cite{Okon2011,Gravitation}. We will investigate the gravitation energy problem from many different aspects: the gravitational $4$-momentum, the stress-energy tensor, and the problem of localizing the energy. We also investigate the impact that the quantization of the energy of the electromagnetic field may have on the spacetime geometry. In doing so, we specialize to the $pp$-waves that are solutions of the Einstein-Maxwell equations.  We show that both the cross-sectional area and the volume must satisfy some constraints for consistency with the quantization of the electromagnetic energy.

The first attempt to find the gravitational stress-energy tensor was done by Einstein in 1915 \cite{TheBerlinYearsVolume6p98}. His definition was considered unsatisfactory, because it was not a tensor under coordinate transformations. This started a quest for the right approach to the gravitation energy, which ended up giving a zoo of pseudotensors \cite{10.2307/20488488,PhysRev.89.400,Weinberg:1972kfs,Landaufourthv2,OZKURT2017}, all of which suffer from the same disease, coordinate dependence. Eventually, the belief that the principle of equivalence prevents one from localizing the gravitational energy spread, and most physicist interpreted this belief as being equivalent to the impossibility of having a well-defined stress tensor (See, e.g, Sec.~20.4 of Ref.~\cite{Gravitation}).

As a result, the focus was shifted towards the definition of the total spacetime energy-momentum, such as the ADM approach \cite{Arnowitt2008}, and  quasilocal definitions \cite{PhysRevD.47.1407,PhysRevLett.83.1897}, which define the spacetime $4$-momentum as a surface term. But those definitions are not considered satisfactory. The ADM is considered too limited, since it is well established only for asymptotically flat spacetimes, and the quasilocal approaches are ambiguous.

An interesting approach was initiated by M\o ller in 1961 \cite{MOLLER1961118,MOLLER1961Mat}. In this approach, M\o ller uses the teleparallel theory that Einstein developed between 1928 and 1931 \cite{SAUER2006399} to deal with the problem of the gravitational energy. Although his approach was not successful, it gave a complete different way of tackling the problem, namely, using the tetrad formalism to remove the coordinate dependence.  

The teleparallel theory invented by Einstein was based on a non-Riemannian geometry known as Weitzenb\"{o}ck geometry. Einstein's goal was to unify gravity with electromagnetism, but he soon realized that was not possible in this geometry. Teleparallelism was revived many times and has been used for many different purpose: dealing with the gravitational energy problem \cite{ANDP:ANDP201200272,Emtsova2021}, formulating the gauge theory of the translation group \cite{PhysRevD.14.2521,HAYASHI1977441,aldrovandi2012teleparallel}, applications to cosmology \cite{PhysRevD.75.084031,PhysRevD.79.124019,PhysRevD.85.124007,Cai_2016}; it has been used by Mashhoon in his nonlocal general relativity \cite{galaxies3010001,Mashhoonbook2017} etc. However, here we focus only on its role in solving the energy problem.

In the case of the energy problem, some approaches to motivate the gravitational energy has been adopted within teleparallelism \cite{ANDP:ANDP201200272,NesterPositive1989,NesterSTGR1999}. Here, we will consider Maluf and collaborators approach \cite{doi:10.1063/1.530774,Maluf1999,PhysRevD.64.084014,PhysRevD.65.124001,PhysRevD.82.124035}. In this approach, Maluf et. al use the teleparallel theory that is equivalent to general relativity (TEGR for short) to define not only the spacetime and gravitational $4$-momenta, but also the stress tensors. In doing so, they use the Hamiltonian formulation of the TEGR. The main difference between this Hamiltonian approach and that of Nester (see, e.g., Ref.~\cite{NesterPositive1989})  is that Maluf really treats the gravitation field as being represented by the tetrad field, while Nester removes the tangent-space index in the definition of the $4$-momentum, which in turn changes the motivation for the definition of the stress tensors (see Sec.~VI of Ref.\cite{PhysRevD.106.044021} for a brief discussion about that point). Furthermore, the $4$-momentum and the gravitational stress-energy tensor of this approach gives consistent results \cite{Gonalves2021,doi:10.1142/S021773232150125X,doi:10.1142/S0217732322502224}, and generalizes the $4$-momentum of the ADM approach \cite{PhysRevD.106.044021}.

Quantum gravity started in the 1930s \cite{https://doi.org/10.48550/arxiv.gr-qc/0006061v3} and, from there on, numerous proposals towards a quantization of gravity have been made. There are so many of them that we can go from theories with many extra dimensions, such as string theory, to theories where the spacetime geometry is  discrete, such as loop quantum gravity. Here we use a semi-classical approach to argue in favor of theories with discrete spacetime geometries. In doing so, we use solutions of the $pp$-wave spacetimes to show that the quantization of the electromagnetic energy leads to a kind of quantization of the area, and also the volume. We discuss the real meaning of this quantization in the context of our semi-classical approach and argue that, if all matter fields are quantized, then the discreteness of geometry should be a fundamental property of spacetime.  

In the next section  we review the TEGR and present two types of irreducible decompositions, one to study the congruence of the observers' trajectories and other to obtain the energy densities, pressures and heat flows of the  stress-energy tensors. We specialize to the $pp$-wave spacetimes with plus polarization in  Sec.~\ref{30012023a}, where general results are obtained and discussed. Two applications are given in Sec.~\ref{20012023n}. We first apply our analysis to an electromagnetic pulse in a spacetime that is asymptotically flat along the $z$ direction. Then we perform the same analysis to the case of a sinusoidal electromagnetic wave oscillating in phase with the metric components. We also show that by a suitable choice of the electric field, Einstein equations become the modified Mathieu equation.  Concluding remarks are made in Sec.~\ref{30012023b}.

\section{The TEGR}
There are many different theories that can be called ``teleparallel theory'' \cite{ANDP:ANDP201200272,NesterSTGR1999,PhysRevD.98.044048,PhysRevD.101.103507,Bahamonde_2023}. This is so because the concept of parallelism depends on the concept of affine connection, and there are many types of connections. The necessary and sufficient condition for teleparallelism was established by Eisenhart in 1927 \cite{Eisenhart1927}. It basically states that any affine connection with vanishing curvature has parallelism, i.e., a set of $n$ linearly independent fields of parallel vectors. As a result, any theory based on a $n$-dimensional manifold endowed with a connection $\nablab$ that has vanish curvature can be called teleparallel. Neglecting the trivial case (Minkowski spacetime), we can separate these theories in three major classes: theories with torsion and vanishing nonmetricity (those are the most common ones) \cite{ANDP:ANDP201200272,PhysRevD.19.3524,Cai_2016}; theories with nonmetricity and vanishing torsion \cite{NesterSTGR1999,PhysRevD.98.044048,PhysRevD.101.103507}; theories with both torsion and nonmetricity\footnote{As far as we know, this kind of theory is not very common in the literature. A limited version of this type of teleparallelism can be found in Ref.~\cite{PhysRevD.99.064047}. In this reference, only the cases in which the Weyl $1$-form is integrable represent teleparallel theories.}. 

We believe that the best teleparallel theory to deal with the problem of describing the gravitational energy is the TEGR. This theory is endowed with two affine connections, the Levi-Civita and Weitzenb\"{o}ck\footnote{Roland Weitzenb\"{o}ck is considered to be the first one to have arrived at this connection \cite{RolandW1923}. For a historical review, see Sauer \cite{SAUER2006399}.)} one. The latter is responsible for the teleparallelism, is also a metric connection, and has torsion.

\subsection{The teleparallel frame}
In any spacetime with teleparallelism, there exists a frame where the connection coefficients vanish. We call this frame the {\it teleparallel frame} and denote it by $\e{_a}$, where Latin indices from the beginning of the alphabet run from\footnote{Parentheses are used to avoid confusion with the spacetime indices. However, we do not use them when writing deltas; for example, $\delta^a_0$ is used in place of $\delta^a_{(0)}$.} $(0)$ to $(3)$, while those from the middle of the alphabet run from $1$ to $3$ and represent tangent space indices only when written in the form $(i)$, $(j)$, etc.  The coframe is denoted by $\teta{^a}$, and its relation with the frame is given by the condition $\teta{^a}(\e{_b})=\delta^a_b$. The frame and the coframe are simply called tetrad field, and their components in the coordinate bases $\partial_\mu$ and $dx^\mu$ are $\e{_a^\mu}$ and $\e{^a_\mu}$, respectively; mathematically, we have $\e{_a}=\e{_a^\mu}\partial_\mu$ and $\teta{^a}=\e{^a_\mu}dx^\mu$, where the Greek indices run from $0$ to $3$. 

In a teleparallel theory with a Weitzenb\"{o}ck connection, one can choose the teleparallel frame as a set of four orthonormal vector fields, i.e, we can choose $\e{_a}$ in such a way that the scalar product satisfies the relation $g(\e{_a},\e{_b})=\eta_{ab}$, where $\eta_{ab}=\textrm{diag}(-1,+1,+1,+1)$ is the Minkowski metric. In terms of components, the orthonormality condition can be expressed as $\e{_a^\mu}\e{_b_\mu}=\eta_{ab}$.

If we write the torsion tensor in this frame, we will find that the components take the form
\begin{align}
\torsion{^a_\mu_\nu}=\pd{_\mu}\e{^a_\nu}-\pd{_\nu}\e{^a_\mu}, \label{04102019p}
\end{align}
which coincide with the object of anholonomity. From these components, one defines the torsion scalar as $T\equiv\potential{^a^b^c}\torsion{_a_b_c}$, where
\begin{equation}
\potential{^\lambda^\mu^\nu}\equiv\frac{1}{4}\left( \torsion{^\lambda^\mu^\nu}+2\torsion{^{[\mu|}^\lambda^{|\nu]}}\right)+g^{\lambda[\nu}T^{\mu]}
\label{10112017l}
\end{equation} 
is the {\it superpotential}. (In the expression above, we are using $\torsion{^\lambda_\mu_\nu}= \e{_a^\lambda}\torsion{^a_\mu_\nu}$ and the definition $\torsion{^\mu}\equiv \torsion{^\lambda_\lambda^\mu}$.) The Lagrangian density is defined as ${\cal L}=-keT-{\cal L}_M$, where ${\cal L}_M$ is the matter Lagrangian density, and $k\equiv c^4/(16\pi G)$ ($c$ and $G$ are the speed of light and the gravitational constant, respectively);  the quantity $e$ stands for the determinant of the tetrad field $\e{^a_\mu}$.\footnote{ When the triad $\e{_{(i)}}$ is right-handed and $\e{_{(0)}}$ points toward the future, we can use $e=+\sqrt{-\det g}$.}  From ${\cal L}$ one constructs the action and obtains the field equations by taking variations with respect to the tetrad field.

\subsection{Field equations}
In GR, Einstein's equations are written in the form 
\begin{equation}
G^{\mu\nu}=\frac{1}{2k}{\cal T}^{\mu\nu},
\label{02012022c}
\end{equation}
where ${\cal T}^{\mu \nu}$ is the matter stress-energy tensor.

The field equations of the TEGR coincide with those of GR, but they are written in such a way that shows explicitly the gravitational energy content. By using the Hamiltonian formulation of the TEGR, Maluf and collaborators defined a $4$-momentum $P^a$ for the spacetime, and  were led to writing  Einstein's equations in the alternative form
\begin{equation}
\pd{_\alpha}\left( e\potential{^a^\mu^\alpha}\right)=\frac{e}{4k}\left(\energy{^\mu^a}+{\cal T}^{\mu a}\right), \label{29032019k}
\end{equation}
where the quantity $\energy{^\mu^a}$ is given by
\begin{equation}
\energy{^\mu ^a}=k(4\potential{^b^c^\mu}\torsion{_b_c^a}-\e{^a^\mu}T) \label{29032019h}
\end{equation}
and is interpreted as the stress-energy tensor of the gravitational field. 

Since the Weitzenb\"{o}ck connection coefficients in the teleparallel basis $\e{_a}$ vanish, one can recast Eqs.~\eqref{10112017l} and \eqref{29032019h} solely in terms of the Levi-Civita connection coefficients. Let $\leviconnection{}$ and $\rsconnection{^a_b_c}$ represent the Levi-Civita connection and its coefficients in the tetrad basis\footnote{These coefficients are frequently called ``spin connection''.}, respectively. Our definition for these coefficients are\footnote{It is also common in the literature to use the notation $\rsconnection{^a_b_c}\equiv \bracket{\teta{^a},\leviconnection{_b} e_c }$ to indicate the action of the $1$-form $\teta{^a}$ on $\leviconnection{_b} e_c$.} $\rsconnection{^a_b_c}\equiv \teta{^a}\left(\leviconnection{_b} e_c\right)$. The superpotential and the gravitational stress-energy can be rewritten as (see, e.g., section 3.5.1 of Ref.~\cite{formiga2022meaning}; see also Ref.~\cite{PhysRevD.73.124017})
\begin{align}
\potential{_a_b_c}=\frac{1}{2}\rsconnection{_c_a_b}+\rsconnection{^d_d_{[c}}\tensor{\eta}{_{b]a}},
\end{align}
\begin{align}
\energy{^b_a}= 2k\Bigl(2\rsconnection{^c_{[ad]}}\rsconnection{^b_c^d}-2\rsconnection{^b_{[ad]}}\rsconnection{^c_c^d}
-\rsconnection{^c_c_a}\rsconnection{^d_d^b}
\nonumber\\
+\delta^{b}_{a}\rsconnection{^c_{[c|f}}\rsconnection{^d_{|d]}^f}  \Bigr),
\end{align}
where $\rsconnection{^a_b_c}$ can be written in terms of the Weitzenb\"{o}ck torsion as
\begin{eqnarray}
\rsconnection{^a_b_c}=\frac{1}{2}\left( \torsion{_b_c^a}+\torsion{_c_b^a}-\torsion{^a_b_c} \right).
\label{10092020a}
\end{eqnarray}
Equation \eqref{10092020a} holds only if the torsion tensor is written in the teleparallel frame.

These expressions are particularly good for studying the relation between the chosen teleparallel frame and the predicted gravitational energy; they are also useful when comparing with GR. For instance, it has been shown that $\energy{^b_a}$ vanishes along an observer worldline regardless of its acceleration if the teleparallel frame is the observer's proper reference frame (theorem 2 of Ref.~\cite{doi:10.1142/S0217732322502224}). 

This theorem may be seen as an indication that the gravitational energy is nonlocal. And the advantage of this result over the vanishing of pseudotensors only in local Lorentz frames is that it does not depend on the observers motion. In this sense, the gravitational stress-energy tensor of the TEGR is more compatible with the modern view, due to Synge, that the gravitational interaction is a consequence of the Levi-Civita curvature, and not a consequence of inertia (or acceleration), as the pseudotensors suggest.

There are many other advantages of $\energy{^\mu ^a}$ over the so-called pseudotensors \cite{PhysRevD.106.044021}. As an example, we can see from Eq.~\eqref{29032019h} and the definition of $T$ that $\energy{^\mu ^a}$ is traceless, which is compatible with the fact that it is supposed to represent the stress-energy tensor of a massless field.  As far as we know, the other approaches, such as Einstein's and Landau-Lifshitz's pseudotensors [see, e.g., Eqs.(20b) and (7.75) of Refs.~\cite{TheBerlinYearsVolume6p98,bambi2018} , respectively], do not yield a traceless pseudotensor.

\subsection{$4$-momentum}
Following the TEGR approach, we interpret $\energy{^\mu ^a}$ as being the stress-energy tensor of the gravitational field, and $\tau^{\mu\nu}\equiv \energy{^\mu^\nu}+{\cal T}^{\mu\nu}$ as the spacetime one. The spacetime, gravitational, a matter $4$-momenta are defined as
\begin{align}
P^a=\int_V d^3x e\tau^{0a},
\label{02012023a}
\end{align}
\begin{align}
P^a_g=\int_V d^3x e\energy{^0^a},\quad P^a_M\equiv\int_V d^3x e{\cal T}^{0a}.
\label{02012023b}
\end{align}
The integrals are over a region $V$, a hypersurface defined by $x^0=$ constant.

If $e\tau^{0a}$ is not singular in the region $V$, we can use Stokes' theorem in Eq.~\eqref{29032019k} and the identity $\potential{^a^0^0}=0$ to recast Eq.~\eqref{02012023a} as a surface integral. The result is
\begin{align}
P^a=4k\oint_S dS_i e\Sigma^{a0i},
\end{align}
where $S$ is the boundary of $V$.

In particular, if the spacetime is such that the metric becomes that of Minkowski when a certain radial coordinate $r$ goes to infinity, then the spacetime total $4$-momentum can be given by
\begin{align}
P^a_\textrm{\tiny total}=4k\lim_{r\to\infty}\oint_{S_e} dS_i e\Sigma^{a0i},
\label{03012023a}
\end{align}
where the integral is calculated over the external boundary $S_e$. If there is no singularity at all, then $S_e=S$ and $P^a_\textrm{\tiny total}$ is just $\lim_{r\to\infty}P^a$.  

However, in case there are singularities, to keep the region $V$ free from singularities, we surround all of them with inner boundaries and rewrite the equation  $P^a=P^a_g+P^a_M$ as\footnote{We have used $P^a=P^a_e+P^a_I$.} $P^a_e=P^a_g+P^a_M-P^a_I$, where $P^a_I=4k\oint_{S_I} dS_i e\Sigma^{a0i}$ is the integral over the inner boundaries, and $P^a_g$ and $P^a_M$ are calculated in the region $V$. Of course, in some cases $P^a_g$, $P^a_M$, and $P^a_I$ may be divergent. But in an asymptotically flat spacetime, one expects $P^a_\textrm{\tiny total}\equiv \lim_{r\to\infty}P^a_e$ to converge for a proper choice of the teleparallel frame.

To finish this section, we point out that it has recently been shown that Eq.~\eqref{03012023a} generalizes the ADM $4$-momentum,  Eqs.~(5.1)-(5.2) of Ref.~\cite{Arnowitt2008}. (See Ref.~\cite{PhysRevD.106.044021}, for proof of this assertion.) This and the fact that the TEGR yields a stress-energy tensor with interesting properties, such as being traceless, justify taking this theory seriously when working with the gravitational energy problem.

\subsection{Irreducible decomposition of $\rcd{_\beta}\e{_{(0)}}$} 
Since both $\tau^{\mu\nu}$ and $\energy{^\mu^\nu}$ depend on the teleparallel frame we choose to work with, different frames give different results, and some of these results are clearly meaningless \cite{doi:10.1142/S0217732322502224}. There are some attempts to  circumvent this problem, but they all seem to fail \cite{Maluf_2020}. This means that the interpretation of $\tau^{\mu\nu}$ and $\energy{^\mu^\nu}$ as the spacetime and gravitational stress-energy tensors is still an open question. Because of that, studying the relation between the teleparallel frame and the predictions made by $\tau^{\mu\nu}$ and $\energy{^\mu^\nu}$ is important to the understanding of the gravitational energy problem.

In studying this relation, it is interesting to understand the properties of the observer congruence. We can analyze a congruence of curves whose tangent vector field is a $4$-velocity field $u$ by decomposing $\rcd{_\beta} u_\alpha$ into its irreducible parts with respect to the rotation group \cite{Ehlers1993,Ellis2009}. (Keep in mind that, in the language we use in this paper, the expression $\rcd{_\beta} u_\alpha$ represents the components of the covariant derivative; it can be obtained from the scalar product between $\partial_\alpha$ and $ \leviconnection{_\beta}u $.)  This decomposition takes the form
\begin{align}
\rcd{_\beta} u_\alpha=\omega_{\alpha\beta}+\theta_{\alpha\beta}-\frac{1}{c^2}a_\alpha u_\beta,
\label{27112022a}
\\
\theta_{\alpha\beta}\equiv \sigma_{\alpha\beta}+\frac{1}{3}\theta h_{\alpha\beta},
\label{27112022b}
\\
h_{\alpha\beta}=g_{\alpha\beta}+u_\alpha u_\beta/c^2,
\label{28112022d}
\end{align}
where $a_\alpha\equiv u^\beta\rcd{_\beta} u_\alpha$ is the {\it acceleration vector field}, $\omega_{\alpha\beta}\equiv (1/2)(\tensor{h}{^\mu_\beta}\rcd{_\mu} u_\alpha-\tensor{h}{^\mu_\alpha}\rcd{_\mu} u_\beta)$ is the {\it vorticity tensor}, and $\theta_{\alpha\beta}\equiv (1/2)(\tensor{h}{^\mu_\beta}\rcd{_\mu} u_\alpha+\tensor{h}{^\mu_\alpha}\rcd{_\mu} u_\beta)$ is the {\it expansion tensor}. The trace-free tensor $\sigma_{\alpha\beta}$ is the {\it shear tensor}, while the trace of $\theta_{\alpha\beta}$, denoted by  $\theta$, measures the expansion of the congruence. From $\sigma_{\alpha\beta}$ and $\omega_{\alpha\beta}$, one defines $\left[(1/2)\sigma^{\alpha\beta}\sigma_{\alpha\beta}\right]^{1/2}$ and $\left[(1/2)\omega^{\alpha\beta}\omega_{\alpha\beta}\right]^{1/2}$ as the {\it shear} and the {\it vorticity}, respectively.

If the coordinate system $(ct,x^j)$ and the vector field $\e{_{(0)}}$ are such that $\e{_{(0)}}=\partial_0$, then the congruence whose tangent vector field is $u=c\e{_{(0)}}$ corresponds to the worldlines of the particles with constant spatial coordinates, i.e., $x^j=$ constant; furthermore, $t$ is the particles' proper time. We assume that we can treat these particles as a fluid and their worldlines as the fluid flow lines. It follows then that, during a small time interval, the effect of $\theta$ alone is to change a fluid sphere to a similar sphere with a different volume but with the same orientation. In turn, the effect of $\sigma_{\alpha_\beta}$ alone is to distort the sphere without changing the volume. Finally, the vorticity tensor $\omega_{\alpha_\beta}$ produces a rigid rotation about some axis. (See Fig.~1 of Ref.~\cite{Ellis2009} for a geometrical picture of these effects.)

For $u=c\e{_{(0)}}$, we have $\rcd{_\beta} u_\alpha=c\e{_a_\alpha}\e{^b_\beta}\rsconnection{^a_b_{(0)}}$. Then we can recast Eq.~\eqref{27112022a} as
\begin{align}
\rsconnection{_a_b_{(0)}}=\frac{1}{c}\left(\omega_{ab}+\theta_{ab}\right)+\frac{1}{c^2}a_a\delta^0_b,
\label{28112022e}
\\
\theta_{ab}=\sigma_{ab}+\frac{1}{3}\theta h_{ab}.
\label{28112022f}
\end{align}
Note that $h_{ab}=\delta^i_a\delta^i_b=\delta^1_a\delta^1_b+\delta^2_a\delta^2_b+\delta^3_a\delta^3_b$, where we omit the ``parentheses'' in the deltas for convenience, i.e., $\delta_a^{(0)}=\delta_a^{0}$. Since $\theta_{\alpha\beta}u^\alpha=0$ and $\theta_{\alpha\beta}=\theta_{\beta\alpha}$, i.e., $\theta_{(0)b}=\theta_{b(0)}=0$, we only need the spatial part,
\begin{align}
\theta_{(i)(j)}=\frac{c}{2}\left(\rsconnection{_{(i)(j)(0)}}+\rsconnection{_{(j)(i)(0)}}\right),
\label{28112022g}
\end{align}
to calculate $\theta_{ab}$. (Note that $\theta=c\rsconnection{_{(i)(i)(0)}}$.)  The same holds for the vorticity tensor:
\begin{align}
\omega_{(i)(j)}=\frac{c}{2}\left(\rsconnection{_{(i)(j)(0)}}-\rsconnection{_{(j)(i)(0)}}\right).
\label{28112022h}
\end{align}

\subsection{Decomposition of the energy-momentum tensor} 
It is also interesting to decompose the stress-energy tensors to analyze their properties from known quantities such as heat and pressure. Here, we use the decomposition of Refs.~\cite{PhysRev.58.919,Ellis2009}. Since, in general, $\tau^{\mu\nu}$ and $\energy{^\mu^\nu}$ are not necessarily symmetric, we apply this decomposition to the symmetric part\footnote{As pointed out by Bergmann and Thomson \cite{PhysRev.89.400}, there is no need to require that any stress tenor be symmetric, unless a net torque could appear in the absence of external forces. Furthermore, $\energy{^\mu^a}$ corresponds to the components of the canonical energy-momentum  $E_\alpha$ defined in the context of the Metric-Affine Gauge Theory of Gravity (see, e.g., section 5.9 of Ref.~\cite{HEHL19951}); so, the interpretation of its antisymmetric part is similar to that of ordinary canonical energy-momentum tensors. (For the decomposition of its antisymmetric part in the cosmological context, see Ref.~\cite{doi:10.1142/S021773232150125X}) In any case, the stress tensors of the $pp$-waves considered here are symmetric.}. 

The decomposition of these tensors with respect to $\e{_{(0)}}$ can be written as \cite{PhysRev.58.919,Ellis2009}
\begin{align}
\tau^{(\mu\nu)}=\rho\e{_{(0)}^\mu}\e{_{(0)}^\nu}+ph^{\mu\nu}+\frac{2}{c}q^{(\mu}\e{_{(0)}^{\nu)}}-{\cal P}^{\mu\nu},
\label{04122022a}
\end{align}
and 
\begin{align}
\energy{^{(\mu\nu)}}=\rho_g\e{_{(0)}^\mu}\e{_{(0)}^\nu}+p_gh^{\mu\nu}+\frac{2}{c}q_g^{(\mu}\e{_{(0)}^{\nu)}}-{\cal P}_g^{\mu\nu},
\label{04122022b}
\end{align}
where
\begin{align}
\rho\equiv\tau_{(\mu\nu)}\e{_{(0)}^\mu}\e{_{(0)}^\nu},\ p\equiv \frac{1}{3}\tau_{(\mu\nu)}h^{\mu\nu},
\nonumber\\
q^\mu\equiv-c\tensor{h}{_\alpha^\mu}\tau^{(\alpha\beta)}\e{_{(0)\beta}},\ {\cal P}^{\mu\nu}\equiv-\tensor{h}{_\alpha^\mu}\tensor{h}{_\beta^\nu}\tau^{(\alpha\beta)}+ph^{\mu\nu}.
\label{04122022c}
\end{align}
One can easily show that $q_\mu \e{_{(0)}^\mu}=0$, ${\cal P}^\mu_\mu=0$, and ${\cal P}_{\mu\nu}\e{_{(0)}^\nu}=0$. (The definitions of $\rho_g$, $p_g$, $q_g^\mu$, and ${\cal P}_g^{\mu\nu}$ are analogous.)

We assume that we can interpret $\rho$, $p$, $q^\mu$, and ${\cal P}^{\mu\nu}$ respectively as energy density, isotropic pressure, energy flux (heat flow), and anisotropic pressure (viscous stress tensor).

\subsection{Material reference systems}
General relativity (and also TEGR) is seen as a gauge theory and, because of this, one assumes that only gauge invariant quantities can be measured. However, as pointed out by Rovelli in Ref.~\cite{CRovelli1991297}, many authors reject this idea because the gauge in this theory reflects the choice of a particular reference system in which the measurements are made. Rovelli summarizes these two point of views as ``the nonlocal point of view'' and ``the local point of view''. In the former, spacetime points are not a priori distinguishable. On the other hand, in the latter, they are distinguishable. As a result, quantities such as the metric components $g_{\mu\nu}$ are not observable in the former viewpoint, but are observable in the latter. The latter interpretation is necessary if the spacetime manifold is interpreted as being the set of physically determined events, as many textbooks do. Although these two point of views are in apparent contradiction, Rovelli shows that they can be reconciled if we take into account the physical system that is used as the reference frame. 

Rovelli's idea is that spacetime points are a priori indistinguishable, but we can use matter to localize them. So, once a physical frame of reference is established, we have a definition of physical spacetime points, and quantities such as $g_{\mu\nu}$ can be an observable.

To establish such a physical frame without any approximation, we must take into account the energy-momentum tensor of the frame in Einstein's field equations, and also consider the equations that determine the motion of the frame. Nevertheless, in some situations, we can neglect them. Throughout this paper we will neglect only the frame energy-momentum tensor,  but not the frame motion. 

We will assume that the three-dimensional space is filled with particles, each carrying a clock, that can be treated as test particles in the sense that their effects on the background geometry is negligible. However, as pointed out by Rovelli,  the degree of freedom of those particles cannot be neglected, they enter in the definition of the observable quantities. In the case of the TEGR, they will play an important role in the calculation of quantities such as the gravitational energy.

\section{$pp$-wave spacetime}\label{30012023a}
The metric of the $pp$-wave spacetime we are going to work with can be written in the form
\begin{align}
ds^2=-c^2dt^2+f(u)^2dx^2+g(u)^2dy^2+dz^2, \label{26122021a}
\\
u\equiv t-z/c. \label{26122021b}
\end{align}
This metric can be used to describe spacetimes with waves propagating along the $z$ axis and with  plus polarization. These waves can be gravitational waves, electromagnetic waves, or both. Here, however, we will be more interested in the case in which the curvature vanishes when the electromagnetic wave is not present; this case is characterized by $f=g$, but we leave this specialization to Sec.~\ref{28112022a}.

\subsection{Choosing the teleparallel frame} 
We believe that, in analyzing the gravitational energy, it is better to work with a system of freely falling test particle as the constituents of the reference frame. The reason for this is that it is easier to interpret the gravitational energy when non-gravitational interactions are not part of the frame. Furthermore, the frame must also be free from artificial properties (For a discussion about the ideal frame to interpret the predictions of $\energy{^\mu^a}$, see Sec.~V of Ref.~\cite{PhysRevD.106.044021} or Sec.~3 of Ref.~\cite{doi:10.1142/S0217732322502224}.)

It can be easily shown that the coordinate system $x^\mu=(ct,x,y,z)$ is adapted to a system of freely falling particles, i.e, the curves with constant values of $x^i$ represent timelike geodesics. A frame that is adapted to these coordinates, and therefore is a freely falling frame\footnote{Be aware that a freely falling frame is not necessarily a local inertial frame of reference \cite{doi:10.1002/andp.201700175}.}, is 
\begin{align}
\e{_a}=\left(\frac{1}{c}\partial_t,\frac{1}{f}\partial_x,\frac{1}{g}\partial_y, \partial_z \right),
\label{02012022a}
\\
\vartheta^a=\left(cdt,fdx,gdy,dz \right),
\label{02012022b}
\end{align}  
where $\vartheta^a$ represents the coframe. 

In the general way that Eqs.~\eqref{02012022a}-\eqref{02012022b} are written, the frame is not yet free from artificial properties (free from rotations that are neither a consequence of gravity nor the motion of the observers, such as the rotations of the tetrad of the spherical coordinate system). To remove, or at least decrease, the artificiality of the frame, we demand that the frame become a global inertial frame of reference in the absence of gravity. We will come back to this point in Sec.~\ref{02022023a}.

To obtain the Weitzenb\"{o}ck torsion, we substitute the components of Eq.~\eqref{02012022b} into Eq.~\eqref{04102019p}. The result is
\begin{align}
\torsion{^a_b_c}=&\ \frac{1}{c}\frac{f^\prime}{f}\delta^a_1\left(\delta^0_b\delta^1_c-\delta^1_b\delta^0_c+\delta^1_b\delta^3_c-\delta^3_b\delta^1_c \right)
\nonumber\\
& +\frac{1}{c}\frac{g^\prime}{g}\delta^a_2\left(\delta^0_b\delta^2_c-\delta^2_b\delta^0_c+\delta^2_b\delta^3_c-\delta^3_b\delta^2_c \right).
\label{28112022b}
\end{align}
From Eq.~\eqref{10092020a}, we obtain the Levi-Civita ``spin connection''
\begin{align}
\rsconnection{_a_b_c}=\frac{1}{c}\Bigl[\frac{f^\prime}{f}\delta^1_b\left(\delta^1_a\delta^0_c-\delta^0_a\delta^1_c+\delta^3_a\delta^1_c-\delta^1_a\delta^3_c\right)
\nonumber\\
+\frac{g^\prime}{g}\delta^2_b\left(\delta^2_a\delta^0_c-\delta^0_a\delta^2_c+\delta^3_a\delta^2_c-\delta^2_a\delta^3_c\right)\Bigr].
\label{28112022c}
\end{align}
The primes indicate differentiation with respect to $u$.

The relation between the acceleration tensor $\acceleration{_a_b}$ and  $\rsconnection{_a_b_c}$ is $\acceleration{_a^b}=c\rsconnection{^b_{(0)}_a}$. Therefore, from the expression above, we see that $\acceleration{_a^b}=0$. This means that the particles that compose the frame are freely falling and the frame is Fermi-Walker transported along the particles' trajectories (the triad $\e{_{(i)}}$ does not rotate).

We can infer the properties of the congruence from Eqs.~\eqref{28112022g}-\eqref{28112022h}   and \eqref{28112022c}. From Eq.~\eqref{28112022c}, we find that $\rsconnection{_{(i)(j)(0)}}=(1/c)\left[(f^\prime/f)\delta^1_i\delta^1_j+(g^\prime/g)\delta^2_i\delta^2_j \right]$. It is clear that $\rsconnection{_{(i)(j)(0)}}=\rsconnection{_{(j)(i)(0)}}$, which implies that the flow is irrotational [see Eq.~\eqref{28112022h}]. We find that the expansion tensor is just 
\begin{align}
\theta_{(i)(j)}=(f^\prime/f)\delta^1_i\delta^1_j+(g^\prime/g)\delta^2_i\delta^2_j,
\label{29112022a}
\end{align}
leading to the expansion 
\begin{align}
\theta=\left(\ln fg\right)^\prime,
\label{29112022b}
\end{align}
and the shear tensor
\begin{align}
\sigma_{(i)(j)}=\frac{1}{3}\Biggl\{\left[\ln\left(\frac{f^2}{g}\right)\right]^\prime\delta^1_i\delta^1_j+\left[\ln\left(\frac{g^2}{f}\right)\right]^\prime\delta^2_i\delta^2_j
\nonumber\\
-\left[\ln\left(fg\right)\right]^\prime\delta^3_i\delta^3_j\Biggr\}.
\label{29112022c}
\end{align}

It is interesting to note that the determinant of the tetrad field is given by
\begin{equation}
e=fg. \label{26122021f}
\end{equation} 
In other words, the expansion is just the rate of change of $\ln e$.

\subsection{Field equations} 

If we calculate Einstein's tensor in the form $G^{\mu a}$ and substitute it into Eq.~(\ref{02012022c}), we will find that the matter stress-energy tensor satisfies the equation
\begin{align}
{\cal T}^{\mu a}=-\frac{c^2}{8\pi G}\left( \frac{f''}{f}+\frac{g''}{g}\right)\Delta^{\mu a}_{03},
\label{26122021c}
\\
\Delta^{\mu a}_{03}\equiv (\delta^\mu_0+\delta^\mu_3)(\delta^a_0+\delta^a_3).
\label{26122021d}
\end{align}

On the other hand, the gravitational stress-energy tensor is\footnote{This result was first obtained in Ref.~\cite{doi:10.1002/andp.201800320}. See the details of the calculation there.}
\begin{equation}
\energy{^\mu^a}=-\frac{c^2}{4\pi G}\frac{f'g'}{fg}\Delta^{\mu a}_{03}.
\label{26122021e}
\end{equation}

From Eqs.~(\ref{26122021e}) and (\ref{26122021c}), we find that
\begin{equation}
e\tau^{\mu a}=-\frac{c^2}{8\pi G}\left(\frac{\partial^2}{\partial t^2} fg \right)\Delta^{\mu a}_{03}.
\label{02012022d}
\end{equation}
Recall that the quantity $\tau^{\mu a}$ represents the total stress-energy tensor, i.e., $\tau^{\mu a}\equiv\energy{^\mu^a}+{\cal T}^{\mu a}$.

\subsection{Proper area and the spacetime energy}\label{01012023a}
Let us now rewrite Eq.~(\ref{02012022d}) in terms of the proper area measured by a particular observer and obtain the spacetime energy.

Suppose that an observer $\cal{O}$ located at $(x_0,y_0,z_0)$  wishes to calculate the proper distance from two freely falling particles at a certain instant $t$, one located at $(x_p,y_0,z_0)$ and the other at $(x_q,y_0,z_0)$, where $x_q > x_p$.  From Eq.~(\ref{26122021a}) we see that this proper distance is $s_x=f(u_0)\Delta x$, where $u_0\equiv t-z_0/c$ and $\Delta x=x_q-x_p$.

Analogously, the proper distance between two freely falling particles located at $(x_0,y_r,z_0)$ and $(x_0,y_s,z_0)$, with $y_s>y_r$, is given by $s_y=g(u_0)\Delta y$, where $\Delta y=y_s-y_r$.

The observer $\cal{O}$ can then define the proper area 
\begin{align}
A\equiv s_x s_y=f(u_0)g(u_0)\Delta x \Delta y
\label{20012023f}
\end{align}
and, using Eq.~(\ref{26122021f}), recast Eq.~(\ref{02012022d}) as
\begin{equation}
\tau^{\mu a}=-\frac{c^2}{8\pi G A}\frac{\partial^2 A}{\partial t^2}\Delta^{\mu a}_{03}.
\label{02012022e}
\end{equation}
(Since $z_0$ is arbitrary, we omit the fact that $\tau^{\mu a}$ above is evaluated at this location.) Hence, the tensor $\tau^{\mu a}$ depends only on the area $A$.

To evaluate the spacetime energy inside the region $x_p<x<x_q$, $y_r<y<y_s$, and $z_<<z<z_>$,  we can use the identity $\partial^2 fg/\partial t^2=c^2\partial^2 fg/\partial z^2$. Using this identity in Eq.~(\ref{02012022d}) and performing the integration with $\mu=0$, we obtain
\begin{align}
P^a=-\frac{c^4}{8\pi G}\left(\delta^a_0+\delta^a_3\right)
\nonumber\\
\times \Delta x\Delta y\left[\left(\frac{\partial fg}{\partial z} \right)(z_>)-\left(\frac{\partial fg}{\partial z} \right)(z_<) \right].
\label{02012022f}
\end{align}

An observer located at $z$ measures a cross-sectional area given by  $A= f(u)g(u)\Delta x\Delta y$. Consider two observers, one at $z_>$ and the other at $z_<$. These observers can measure the rate of change of the areas $f(t-z_>/c)g(t-z_>/c)\Delta x\Delta y$ and $f(t-z_</c)g(t-z_</c)\Delta x\Delta y$, where the value of $\Delta x\Delta y$ is chosen to be the same for both observers. Writing Eq.~(\ref{02012022f}) in terms of these areas, we obtain
\begin{align}
P^a=-\frac{c^4}{8\pi G}\left(\delta^a_0+\delta^a_3\right)\left(\frac{\partial A}{\partial z}\Big|_{z_>}-\frac{\partial A}{\partial z}\Big|_{z_<}\right).
\label{02012022g}
\end{align} 

In the limit as $z_>$  ($z_<$) goes to $+\infty$ ($-\infty$), the areas  become the proper area measured by the observers far away from the ``source''. In case there is no localization, but the field is periodic, we can take $z_>-z_<$ as representing a wavelength (or a multiple of it). In this last case, the areas in Eq.~\eqref{02012022g} are measured by observers located at opposite faces of a box that is one wavelength long. So, it is clear that, if the rate of change of these areas are the same, then the total spacetime energy in the box is zero. These will be exactly the cases considered in Sec.~\ref{20012023n}.

A comment is in order here regarding the periodic case. In this case, we will find more convenient  to write the final results for $P^a_g$ and $P^a_M$ in terms of the proper area measured by an observer at the center of the box.

\subsection{The gravitational energy density}
The issue of localizing the gravitational energy is a controversial one, mainly because of the principle of equivalence. Most physicists do not believe that we can detect the gravitational field locally (see, e.g., section 20.4 of Ref.~\cite{Gravitation}); there is even a debate over the real meaning of the term ``gravitational field'' (see, e.g., Ref.~\cite{NORTON1985203}; see also section 16.5 of Ref.~\cite{Gravitation}). However, it seems to be clear that we do find ``gravitational entities'' that can be defined at a specific location: the curvature tensor is defined at a point, the metric tensor components at an event does not have to be that of Minkowski, etc; Ohanian has even claimed that gravity can be detected locally \cite{doi:10.1119/1.10744} (see also section 1.7 of Ref.~\cite{OhanianBook}). Furthermore, Rovelli showed how we can localize the spacetime points by treating the frame as a real physical entity \cite{CRovelli1991297}, and not as a mere abstraction. Following these lines, we wish to find an expression for the gravitational energy density at each point of the spacetime manifold in terms of the observer's measurements, regardless of whether this is a local density or just a nonlocal phenomenon due to some composite system (as those discussed in Ref.~\cite{doi:10.1119/1.4895342}). 

We can rewrite Eq.~(\ref{26122021e}) in terms of the proper area $A$ and the proper lengths $s_x$ and $s_y$  measured by an observer located at $z_0$. Recall that these quantities are defined by $s_x\equiv f(u_0)\Delta x$, $s_y\equiv g(u_0)\Delta y$ and $A\equiv s_x s_y$. Using these definitions in Eq.~(\ref{26122021e}), we find that
\begin{equation}
\energy{^\mu^a}=-\frac{c^2}{4\pi GA}\left(\frac{\partial s_x}{\partial t}\right) \left(\frac{\partial s_y}{\partial t}\right) \Delta^{\mu a}_{03}.
\label{13012023a}
\end{equation}

One can argue that, in some sense, this equation realizes the localization of the gravitation energy, thus showing that the principle of equivalence does not prevent us from localizing this energy. The observer can detect this energy by just measuring the proper distances $s_x$ and $s_y$ and their rate.\footnote{Recall that we are taking step 1 of Ref.~\cite{CRovelli1991297}, p. 304, i.e., we are neglecting the stress-energy tensor of the free particles. This is clear in Eq.~\eqref{26122021c}: since ${\cal T}^{\mu a}$ is traceless, it cannot account for massive particles. Hence, in this sense, Eq.~\eqref{13012023a} is an approximation.}

Of course, if we have used a system with non-gravitational interactions as part of the frame, we could obtain a vanishing energy density along the freely falling observer at $z_0$. This is so because gravity could be suppressed by other interactions. The vanishing of $\energy{^\mu^a}$ along the observer's worldline when the teleparallel frame is the observer's proper reference frame (theorem 2 of Ref.~\cite{doi:10.1142/S0217732322502224}) strongly supports that interpretation. However, in a frame with non-gravitational interactions, the meaning of $\energy{^\mu^a}$ as a pure description of the gravitational energy density would be questionable, as discussed in Refs.~\cite{PhysRevD.106.044021,doi:10.1142/S0217732322502224}.

It is interesting to note that the gravitational energy will be negative whenever the rates of change of $s_x$ and $s_y$ have the same sign, and positive when they do not. As we will see, in the pure electromagnetic case, the gravitational energy density is always negative.

\subsection{Decomposition of the energy-momentum tensors} 
Since the components of these tensors are coordinate dependent, we present here the quantities given by Eq.~\eqref{04122022c} multiplied by $e$. From Eqs.~\eqref{04122022c} and \eqref{02012022d}, we find that
\begin{align}
\tilde{\rho}\equiv e\rho=-2\frac{k}{c^2}\left(fg\right)^{\prime\prime},\ \tilde{p}=\frac{1}{3}\tilde{\rho},\ \tilde{q}^\mu=c\tilde{\rho}\delta^\mu_3,
\nonumber\\
\tilde{P}^{ab}=\frac{\tilde{\rho}}{3}\left(\delta^a_1\delta^b_1+\delta^a_2\delta^b_2-2\delta^a_3\delta^b_3\right),
\label{05122022a}
\end{align}
where we have written the viscous stress tensor in the tetrad basis to avoid a possible divergence due to a coordinate singularity.

Now, from Eqs.~\eqref{04122022c}, \eqref{26122021f} and \eqref{26122021e}, we find that
\begin{align}
\tilde{\rho}_g=-4\frac{k}{c^2}f^\prime g^\prime,\ \tilde{p}_g=\frac{1}{3}\tilde{\rho}_g,\ \tilde{q}_g^\mu=c\tilde{\rho}_g\delta^\mu_3,
\nonumber\\
\tilde{P}^{ab}_g=\frac{\tilde{\rho}_g}{3}\left(\delta^a_1\delta^b_1+\delta^a_2\delta^b_2-2\delta^a_3\delta^b_3\right).
\label{05122022b}
\end{align}

For ${\cal T}^{\mu\nu}$, we use Eqs.~\eqref{04122022c} and \eqref{26122021c}. The result is
\begin{align}
\tilde{\rho}_M=-2\frac{k}{c^2}\left(gf^{\prime\prime}+fg^{\prime\prime}\right),\ \tilde{p}_M=\frac{1}{3}\tilde{\rho}_M,\ \tilde{q}_M^\mu=c\tilde{\rho}_M\delta^\mu_3,
\nonumber\\
\tilde{P}^{ab}_M=\frac{\tilde{\rho}_M}{3}\left(\delta^a_1\delta^b_1+\delta^a_2\delta^b_2-2\delta^a_3\delta^b_3\right).
\label{06122022a}
\end{align}
Notice that $\tilde{\rho}=\tilde{\rho}_g+\tilde{\rho}_M$.

As expected, they all satisfy a radiation-like equation of state and the heat flow is in the $z$ direction.

Denoting the temperature, entropy and volume by $T$, $S$ and $V$, respectively, Ulhoa et al. proposed an expression to calculate $T\left(\partial S/\partial V\right)_T$ in the context of the TEGR \cite{doi:10.1142/S0217732322502194}. They applied that expression to $pp$-wave spacetimes using a frame that is not adapted to freely falling particles and obtained a negative result. This result seems to suggest that an isothermal propagation is not allowed.  On the other hand,  one can show that $T\left(\partial S/\partial V\right)_T$ vanishes when the teleparallel frame is given by\footnote{Equation (18) in Ref.~\cite{doi:10.1142/S0217732322502194} can be recast as $T\left(\partial S/\partial V\right)_T=e\left( \tau^{0(0)}-\e{^{(i)}_j}\tensor{\tau}{^j_{(i)}} \right)$. From  Eqs.~\eqref{02012022d} and \eqref{02012022b}, we find that $T\left(\partial S/\partial V\right)_T=0$.} Eq.~\eqref{02012022a}. Therefore, it is possible that the result $T\left(\partial S/\partial V\right)_T<0$ obtained by Ulhoa et al. is a consequence of neglecting the degrees of freedom of the nongravitational fields responsible for the frame accelerations. (Of course, there is always the possibility that the frame possesses some kind of artificial property.)

\subsection{Classical analogue of the number of photons/gravitons} \label{10012022f}
As is well known the relation $E/\omega$, where $E$ is the energy of a photon and  $\omega$ its angular frequency, remains the same for all inertial observers in Minkowski spacetime (see, e.g., section 4.5 of Ref.~\cite{Inverno}; see also Ref.~\cite{JEAvron_1999}). Here, we show that this property remains true for $pp$-waves when the observers are related to each other through a global Lorentz transformation that represents a boost along the $z$-direction.

Let ${\cal P}^a$ represent a quantity that satisfies the properties ${\cal P}^{(0)}={\cal P}^{(3)}$ and ${\cal P}^{(1)}={\cal P}^{(2)}=0$. ( $P^a$, $P_g$, and $P_M$ have these properties.) Consider a new frame $\e[\bar]{_a}$ related to $\e{_a}$ by $\e[\bar]{_a}=\lorentz{_a^b}\e{_b}$, where $\lorentz{_a^b}$ is the boost 

\begin{equation}
\lorentz{_a^b}=\left(\gamma\delta^0_a+\beta\gamma\delta^3_a \right)\delta^b_0+\left(\beta\gamma\delta^0_a+\gamma\delta^3_a \right)\delta^b_3+\delta_a^1\delta^b_1+\delta_a^2\delta^b_2.
\label{03012022a}
\end{equation}
(This is a boost along the $z$-axis.)

Applying the properties of ${\cal P}^a$ to $\bar{{\cal P}}^a=\lorentz{^a_b}{\cal P}^b$, we find that $\bar{{\cal P}}^{(0)}=\sqrt{(1-\beta)/(1+\beta)}{\cal P}^{(0)}$. In turn, if we assume that $\beta$ and $\gamma$ are constant, then the new frame can be written as $\e[\bar]{_{(0)}}=\partial/\partial \bar{t}$, $\e[\bar]{_{(3)}}=\partial/\partial \bar{z}$, $\e[\bar]{_{(1)}}=(1/f)\partial/\partial \bar{x}$, $\e[\bar]{_{(2)}}=(1/g)\partial/\partial \bar{y}$, where $\bar{t}=\gamma\left(t-\beta z \right)$, $\bar{z}=\gamma\left(z-\beta t \right)$, $\bar{x}=x$, and $\bar{y}=y$ are the usual Lorentz transformations. (Note that we are using $c= 1$ here; note also that $\bar{t}$ and  $\bar{z}$ are the proper coordinates of the observers adapted to the new frame.)  It follows from the coordinate changes that $u=t-z=\sqrt{(1-\beta)/(1+\beta)}\bar{u}$, where $\bar{u}\equiv \bar{t}-\bar{z}$.  Therefore, the function $f(u)=F(\omega u)$ will become $f(u)=F(\omega\sqrt{(1-\beta)/(1+\beta)} \bar{u})$. (The same holds for $g(u)$.) This means that the new observers will perceive the angular frequency of the wave as  $\omega\sqrt{(1-\beta)/(1+\beta)}$. In other words, $\omega$ transforms in the same way as the energy. So, we conclude that ${\cal P}^{(0)}/\omega=\textrm{constant}$. 

Generalization of the above result to the case where $\beta$ is not constant may not be possible, and we will not pursue this issue here.

For observers that are related to each other by the above transformations, we can write $E/\omega=C\hbar$, where $\hbar$ is the reduced Planck constant and $C$ is a constant that, in principle, depends on the system. In Sec.~\ref{14022023a}, we will assume that $C$ is related to the number of photons/gravitons and show that this assumption leads to a sort of ``quantization of area.''

\subsection{Maxwell energy-momentum tensor}
The stress-energy tensor of the electromagnetic field can be written in SI units as
\begin{equation}
{\cal T}^{\mu\nu}=\frac{\epsilon_0}{4\pi}\left( F^{\mu\alpha}F^{\nu\beta}g_{\alpha\beta}-\frac{1}{4}g^{\mu\nu}F_{\alpha\beta}F^{\alpha\beta} \right),
\label{03012022b}
\end{equation}
where $F_{\mu\nu}=\partial_\nu A_\mu-\partial_\mu A_\nu$, and we use the $4\pi$ in the denominator for convenience. Here, we deal only with the cases where $A_\mu={\cal A}(u)\delta_\mu^1$, which satisfy Maxwell's equations in the spacetime (\ref{26122021a}) with vanishing current. For this case, we must have
\begin{align}
{\cal T}^{\mu a}=\frac{\epsilon_0}{4\pi c^2}\frac{\left({\cal A}^\prime\right)^2}{f^2}\Delta^{\mu a}_{03}.
\label{03012022c}
\end{align}

The only nonvanishing components of the electromagnetic fields are $E_x={\cal A}^\prime/c$ and $B_y={\cal A}^\prime/c^2$. So, the electromagnetic wave is linearly polarized along the $x$-coordinate direction.

Since $E_x$ depends on the coordinate system, it will be interesting for us to work with $E_{(1)}=F_{(1)(0)}$. The relation between $E_{(1)}$ and $E_x$ is $E_x=fE_{(1)}$.

\section{Pure electromagnetic waves}\label{14022023a} 
In this section we specialize to the case of ``pure\footnote{The word ``pure'' here means that there is no gravitational field (vanishing Levi-Civita curvature) when the electromagnetic wave vanishes.}'' electromagnetic waves, and study the consequences of the hypothesis that the energy of photons is quantized.

\subsection{Electromagnetic wave} \label{28112022a}
In describing a pure electromagnetic wave, we take\footnote{For more details about this choice, see, e.g., section 35.11 of Ref.~\cite{Gravitation} or chapter 4 of Ref.~\cite{griffiths1991colliding}} $g(u)=f(u)$. From Eqs.~(\ref{03012022c}) and (\ref{26122021c}), we see that this choice leads to the equation
\begin{equation}
f^{\prime\prime}=-\frac{\epsilon_0 G}{c^2}E^2_{(1)}f,
\label{090102022a}
\end{equation}
where we have used ${\cal A}^\prime=cfE_{(1)}$. In turn, from Eq.~\eqref{26122021e}, we see that the gravitational stress-energy becomes
\begin{equation}
\energy{^\mu^a}=-\frac{c^2}{4\pi G}\frac{\left(f^\prime \right)^2}{f^2}\Delta^{\mu a}_{03}.
\label{090102022b}
\end{equation}

Note that the gravitational energy is always negative in this case. However, this does  not mean that the spacetime energy is negative everywhere. Since the electromagnetic energy is positive, we may still have regions with positive energy. 

One of the advantages of a  negative gravitational energy is the possibility of having a universe with zero energy, as postulated by Tryon \cite{TRYON1973}. Tryon's idea is that the universe may be just a vacuum fluctuation. The solutions we will be dealing with in the next sections are not supposed to be cosmological models, but the total spacetime energy turned out to be zero, thus allowing us to see them as possible vacuum fluctuations.

\subsection{The energy of photons} 
Here we show that the hypothesis of the quantization of the energy of photons leads to a sort of quantization of the area. Then, we discuss the real meaning of this quantization.

Substituting Eqs.~\eqref{03012022c} and \eqref{090102022b}   into Eq.~\eqref{02012023b}, we find that
\begin{align}
E_\mathrm{M}=\frac{\epsilon_0}{4\pi}\int d^3xE^2_x,
\label{20012023a}
\end{align}
and 
\begin{align}
E_\mathrm{g}=-\frac{c^2}{4\pi G}\int d^3x \left(f^\prime\right)^2,
\label{20012023b}
\end{align}
where we have used Eq.~\eqref{26122021f} and $E_\mathrm{M}\equiv P_\mathrm{M}^{(0)}$ etc. The similarity between these two expressions is astonishing (recall that $E_x={\cal A}^\prime/c$); furthermore, they remind us of energies defined in Minkowski spacetime. It is highly unlikely that one will find similar expressions for the electromagnetic and gravitational energies in a frame that is not a frame of freely falling particles. This suggests that we are on the right track.

To analyze the hypothesis that $E_\mathrm{M}$ is quantized, we need to think about the physically meaningful quantities that are present in the electric field. Since we are dealing with waves, there must exist at least two constants, say  $\omega$ and $E_0$, such that $\omega u$ is dimensionless and $E_0$ is somehow related to the amplitude of the electric field. Since, by definition, $u$ has dimension of time, $\omega$ has units of inverse time (units of frequency). Next, we proceed to write some expressions in terms of $\omega$ and $E_0$ explicitly.

We can, without loss of generality, rewrite $E_{(1)}$ and $f$  as 
\begin{align}
E_{(1)}(u)=E_0F_1(\theta),
\label{20012023c}\\
f(u)=F_2(\theta),
\label{20012023d}
\\
\omega_0^2\equiv\frac{\epsilon_0 G E_0^2}{c^2},
\label{20012023e}
\end{align}
where $F_1(\theta)$ and $F_2(\theta)$ are dimensionless functions to be determined, and $\theta=\omega u$. We assume that $\omega_0$, $\omega$, and $E_0$ are all positive. The meaning of the frequency $\omega_0$ is not clear, but we will use the definition \eqref{20012023e} to recast Eq.~\eqref{090102022a} in a more convenient form. As is clear from $\omega_0\approx 10^{-19}\left[\textrm{m/V$\cdot$ s}\right]E_0$, the value of this frequency is small, unless $E_0$ is huge\footnote{ To get an idea of how small $\omega_0$ is, consider an electric field with the amplitude $100\,$V/m. Assume that we have a periodic wave with angular frequency $\omega_0$. The period associated with this amplitude is of the order of the age of the universe.}.

To calculate the energy, we use the same box discussed in Sec.~\ref{01012023a}. The area $A$, which is given by Eq.~\eqref{20012023f}, can be recast as $A=f(u_0)^2\Delta x\Delta y$ (recall that $u_0=t-z_0/c$, where $z_0$ is the location of the observer at the instant $t$). Using $E_x=fE_{(1)}$ and substituting Eqs.~\eqref{20012023c} and \eqref{20012023d} into \eqref{20012023a} and \eqref{20012023b}, respectively,  we find that
\begin{align}
E_\mathrm{M}=\frac{\alpha_1c^3}{G}A\frac{\omega_0^2}{\omega},
\label{20012023g}
\end{align}
\begin{align}
E_\mathrm{g}=-\frac{\alpha_2c^3}{G}A\omega,
\label{20012023h}
\end{align}
where $\alpha_1$ and $\alpha_2$ are dimensionless constants given by
\begin{align}
\alpha_1\equiv\frac{I_1}{4\pi f(u_0)^2},\quad \alpha_2=\frac{I_2}{I_1}\alpha_1,
\label{20012023i}
\end{align}
and $I_1$ and $I_2$ are given by the integrals
\begin{align}
I_1\equiv-\int_{\theta(z_<)}^{\theta(z_>)}d\theta F_1(\theta)^2F_2(\theta)^2,
\label{20012023j}
\end{align}
\begin{align}
I_2\equiv-\int_{\theta(z_<)}^{\theta(z_>)}d\theta  \left[\frac{d }{d\theta}F_2(\theta)\right]^2.
\label{20012023l}
\end{align}

The values of $z_<$ and $z_>$ is chosen in a convenient way. In the examples of Sec.~\ref{20012023n}, we will use $z_<\to-\infty$ and $z_>\to\infty$ for the asymptotically flat case, and $z_>-z_<=n\lambda$ (a natural number times the wavelength) for the periodic one.

Let us assume that the dependence of $F_2$ on $\omega$ is of the form  $F_2(\theta)=F_3(\omega)F(\theta)$, where $F(\theta)$ depends on $\omega$ only through $\theta=\omega u$. In this case, the constant $\alpha_2$ will not depend on $F_3$ because of the term $I_2/f^2(u_0)$ present in  $\alpha_2$. Furthermore, if both  $\theta(z_<)$ and $\theta(z_>)$ are such that $I_2$ do not depend on $\omega$, then $\alpha_2$ will be completely independent of $\omega$.  Therefore, for  this type of solution, the gravitational energy will be proportional to $\omega$ and will resemble the energy of particles in quantum mechanics.

On the other hand, the assumption that  $\alpha_1$ does not depend on $\omega$ ensures that the electromagnetic energy is not exactly proportional to $\omega$ if $\omega\neq\omega_0$. Nevertheless, for the cases in which $\alpha_1\propto\omega^2/\omega_0^2$, the energy $E_\mathrm{M}$ will also resemble $\hbar\omega$.

As shown in section \ref{10012022f}, the relation $P^{(0)}_M/\omega$ is a constant for all sets of freely falling observers that are related to each other through a boost along the $z$-axis with constant velocity. We do not know the value of this constant. But, if this constant were derived from quantum mechanics, its value would probably be a function of the number of photons. If the electromagnetic energy inside the box is quantized in the same fashion as in Minkowski spacetime, then it is natural to assume that $E_\mathrm{M}=C(N)\hbar\omega$, where $C(N)$ is a function of the number of photons $N$ inside the box. (In principle, $C$ can only be determined from a quantum theory of gravity.) From this assumption and Eqs.~\eqref{20012023g} and \eqref{20012023e}, we obtain
\begin{align}
E_0=\sqrt{\frac{\hbar}{\epsilon_0c}\frac{C(N)}{\alpha_1 A}}\omega.
\label{20012023m}
\end{align}
Since $E_0$ is a constant, any change in the value of $A$ must be followed by an equal change in $C(N)$. Of course, if we increase the size of the box, the number of photons inside the box will also increase. The problem, however, is that $C(N)$ cannot change continuously. This means that the expression above is well defined only for discrete changes of $A$. To be more precise, we must have $A\propto C(N)$. Another way of seeing this is to note that $C(N)/A$ must be a constant. So, if we demand that Eq.~\eqref{20012023m} be always well defined, then we would have to use only these values for the area.

The ``quantization'' given by Eq.~\eqref{20012023m} is not a fundamental property of the spacetime. It is possible that it is saying that, for the system we are considering here, we cannot find a certain number of photons inside a box that has an arbitrary small area in the $xy$ plane. For instance, for one photon, we could say that this photon ``occupy'' a minimal area given by $N=1$ (assuming that $C$ increases with $N$); for the ground state, the minimal area for this system to have a measurable energy would be given by $N=0$. But the spacetime may still allow for the measurement of smaller areas by using the test particles of the reference frame, which are not quantized here. 

However, this result is suggesting an interesting possibility: if we quantize all non-gravitational fields, including those that are the constituents of the frame, we might end up with a quantum theory for gravity. The metric would be a quantized field because it comes from the frame, and the frame is adapted to some matter field, which is supposed to be quantized. This means that we should identify the physical system behind $\e{_a^\mu}$ and then quantize it. The result would be a quantum frame\footnote{The idea of a quantum frame of reference is not a new one. See, for example, Rovelli's argument for taking the bodies used in the system of reference seriously \cite{CRovelli1991297}. For more ideas about this issue, see also  Refs.~\cite{CRovelli1991,PhysRevD.30.368}.} and a possible consistent theory for quantum gravity.

The reader might not like the idea of a photon ``occupying a certain region'' referred to above. Nevertheless, by following quantum philosophy, one might avoid this idea by exchanging it for ``having a well-defined value''. For instance, if the observer measures an area that does not satisfy Eq.~\eqref{20012023m}, then one could say that the number of photons inside the box will be undetermined. In this case, it is not clear whether the spacetime would be the same. For consistency, it would probably be slightly different.

The idea that $A$  has to be discrete, for a well-defined number of photons inside the box, will become much clearer in the following sections.

\subsection{Quantum hypothesis for the gravitational energy}
Let us assume that the gravitational energy is also quantized and can be given by $|E_g|=C_\mathrm{g}(N_\mathrm{g})\hbar\omega$, where $N_\mathrm{g}$ is the number of gravitons inside the box. (Like the electromagnetic case, $C_\mathrm{g}$ can be determined only from a quantum theory of gravity.) Applying this assumption to Eq.~\eqref{20012023h}, we obtain
\begin{align}
A=\frac{C_\mathrm{g}(N_\mathrm{g})}{\alpha_2}l_p^2,
\label{20012023o}
\end{align}
where $l_p\equiv \sqrt{\hbar G/c^3}$ is the Planck length. We have again a sort of ``quantization of the area''. 

For consistency, if the observer measures the gravitational and the electromagnetic energies simultaneously, and the number of photons and gravitons are well defined, then both Eqs.~\eqref{20012023m} and \eqref{20012023o} must also hold simultaneously. This leads to 
\begin{align}
\frac{I_1}{I_2}=\frac{C(N)}{C_\mathrm{g}(N_\mathrm{g})}\frac{\omega^2}{\omega_0^2}.
\label{20012023p}
\end{align}

This constraint is identically satisfied for spacetimes with vanishing energy. To see this, consider Eqs.~\eqref{20012023g}-\eqref{20012023h} and their quantum versions. Since the spacetime energy is just $E=E_\mathrm{g}+E_\mathrm{M}$, we have $E=-\alpha_2 c^3 A\omega/G+\alpha_1 c^3 A\omega_0^2/(G\omega)$ and $E=-C_\mathrm{g}(N_\mathrm{g})\hbar\omega+C(N)\hbar\omega$. Taking $E=0$ in both expressions, we obtain $\alpha_1/\alpha_2=\omega^2/\omega_0^2$ and $C_\mathrm{g}(N_\mathrm{g})=C(N)$. Using these expressions and Eq.~\eqref{20012023i}, one sees that Eq.~\eqref{20012023p} is satisfied. 

\subsection{Spacetimes with vanishing energy}
Let us now assume that we do not know whether gravitons exist, but we known that $E_\mathrm{M}=C(N)\hbar\omega$ and $E=0$. In this case, we must have $E_\mathrm{g}=-E_\mathrm{M}=-C(N)\hbar\omega$, i.e., the gravitational energy is also quantized. As a result, we find that the area must satisfy 
\begin{align}
A=\frac{C(N)}{\alpha_2}l_p^2
\label{04022023a}
\end{align}
for consistency.

We will explore the consequences of Eq.~\eqref{04022023a} in the examples of Sec.~\ref{20012023n}.

\subsection{Energy densities}\label{26012023d} 
In order to rewrite Eqs.~\eqref{03012022c} and \eqref{090102022b} in terms of $F_1$ and $F_2$, we use Eqs.~\eqref{20012023c}-\eqref{20012023e}. The result is
\begin{align}
e{\cal T}^{\mu a}=\tilde{\rho}_0 F_2(\theta)^2F_1(\theta)^2\Delta^{\mu a}_{03},
\label{26012023a}
\end{align}
\begin{align}
e\energy{^\mu^a}=-\frac{c^2\omega^2}{4\pi G}\left[\frac{d}{d\theta} F_2(\theta)\right]^2\Delta^{\mu a}_{03},
\label{26012023b}
\end{align}
where
\begin{align}
\tilde{\rho}_0=\frac{\epsilon_0 E_0^2}{4\pi},
\label{26012023c}
\end{align}
and we have multiplied the expressions by $e=f^2$; we have also used the fact that ${\cal A}^\prime=cfE_{(1)}$.

\subsection{Field equations}
Now we rewrite the field equations  in terms of $F_1$ and $F_2$. Substitution of Eqs.~\eqref{20012023c}-\eqref{20012023d} into  Eq.~\eqref{090102022a}  gives
\begin{align}
\omega^2\frac{d^2}{d\theta^2}F_2(\theta)=-\omega_0^2F_1(\theta)^2F_2(\theta).
\label{21012023a}
\end{align}

This equation will be used to find some particular solutions. In what follows, we use it to analyze the limit where gravity is absent. 

\subsection{Absence of gravity}\label{02022023a}
We demand that  the teleparallel frame should become holonomic (a global inertial frame of reference)  when the curvature of the Levi-Civita connection vanishes (absence of gravity). A necessary and sufficient condition for this to hold is that $\rsconnection{^a_b_c}$ vanishes (see, e.g., pages 14 and 15 of  Ref.~\cite{doi:10.1142/S0217732322502224}) in the absence of gravity.

It is clear from Eq.~\eqref{28112022c} that the connection coefficients $\rsconnection{^a_b_c}$ are all proportional to $f^\prime/f$, where we are using  $g=f$ here. In turn, the absence of gravity can be defined by the limit $\omega_0\to 0$ (Recall that $\omega_0\propto E_0$). Therefore, we demand that $\lim_{\omega_0 \to 0}f^\prime/f=0$.

It is more convenient, though, to write this limit in terms of $\omega$. From Eq.~\eqref{21012023a}, we see that $\omega^2 d^2F_2/d\theta^2=0$ when $\omega_0=0$. The solution for $\omega$ different from zero yields $F_2=C_1u+C_2$, where $C_1\propto \omega$. Since $f(u)=F_2(\omega u)$, we have $f(u)=C_1u+C_2$, which leads to $f^\prime/f=C_1/(C_1u+C_2)$. It is clear that the frame will not be holonomic unless $\omega=0$ (the same as $C_1=0$) and $C_2\neq 0$. So, in the absence of gravity, we demand that
\begin{align}
\lim_{\omega\to 0}\frac{f^\prime}{f}=0.
\label{28012023a}
\end{align} 
This condition is independent of the coordinate system, since the coefficients $\rsconnection{^a_b_c}$ are scalars under coordinate transformations. 

Due to technical difficulties, we relax this condition in the asymptotically flat example of the next section.

\section{Worked examples}\label{20012023n} 
In this section we will analyze the spacetime, gravitational, and electromagnetic energies of two special cases: a electromagnetic wave that is asymptotically flat along the $z$-axis, and a ``pure'' sinusoidal electromagnetic wave.

\subsection{An asymptotically flat case}\label{01012023b}
By asymptotically flat, here, we mean that the metric becomes the Minkowski metric when $u$ goes to $\pm \infty$ (or, equivalently, $z\to \mp\infty$ with $t$ fixed). This limit represents a region far away from the source, which will be localized in the neighborhood of $u=0$.

To ensure that the spacetime is asymptotically flat, we first choose $F_2$ and only then search for the $F_1$ that satisfies Eq.~\eqref{21012023a}.

\begin{figure}
\includegraphics[scale=0.3]{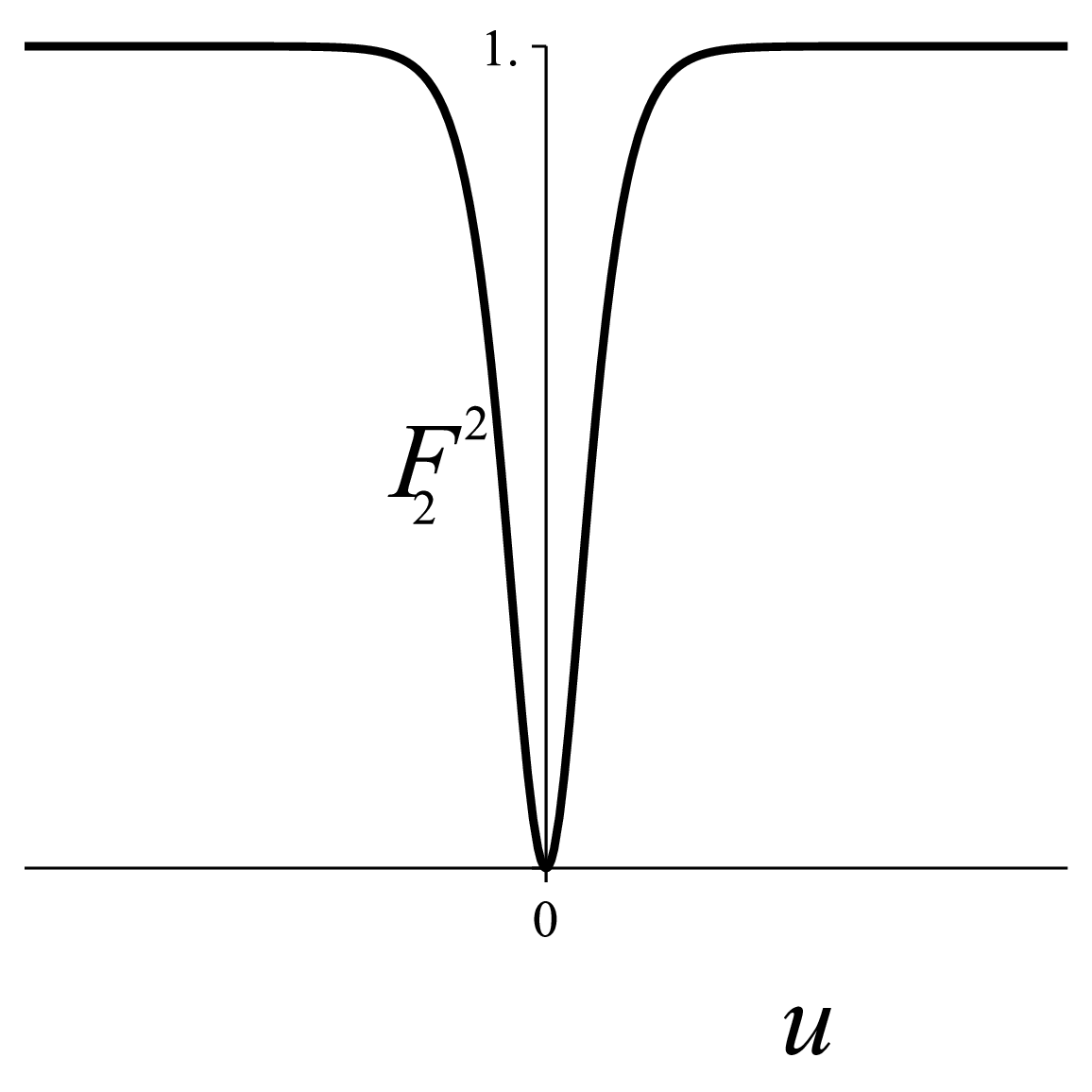}
\caption{The behavior of $F_2^2$ as a function of $u$. We have used $\omega=1$.}
\label{29112022d}
\end{figure}

Let us search for a solution with 
\begin{equation}
F_2=\tanh\left(\omega u\right).
\label{10012022a}
\end{equation}
This choice describes a pulse in the neighborhood of $u=0$ moving along the $z$-axis in the positive direction (see Fig.~\ref{29112022d}). Unfortunately, this choice does not lead to a holonomic frame in the limit $\omega\to 0$, because the constant $C_2$ defined in Sec.~\ref{02022023a} is zero. However, we will use it as an example  because the frame becomes holonomic in the absence of gravity when $u$ goes to $\pm\infty$ and, as we will see, the densities will vanish in the absence of gravity.   

Substitution of Eq.~\eqref{10012022a} into Eq.~\eqref{21012023a} yields $F_1=(\sqrt{2}\omega/\omega_0)\sech \theta$. There is a subtlety in this result: substituting it in Eq.~\eqref{20012023c} and using \eqref{20012023e}, we see that $E_0$ will disappear, and $\omega$ will become the ``amplitude''. This means that $\omega$ and $E_0$ are not independent parameters. We can, however, choose the relation between them. If we want $F_1$ to be just $\sech \theta$, we must take $\sqrt{2}\omega/\omega_0=1$. So, we have
\begin{align}
F_1=\sech(\omega u),
\label{21012023c}
\end{align}
and
\begin{align}
\omega=\frac{\omega_0}{\sqrt{2}},\quad (E_0=c\omega\sqrt{\frac{2}{\epsilon_0 G}}).
\label{06022023a}
\end{align}
In other words, there is only one free parameter in this solution.

\begin{figure}
\includegraphics[scale=0.3]{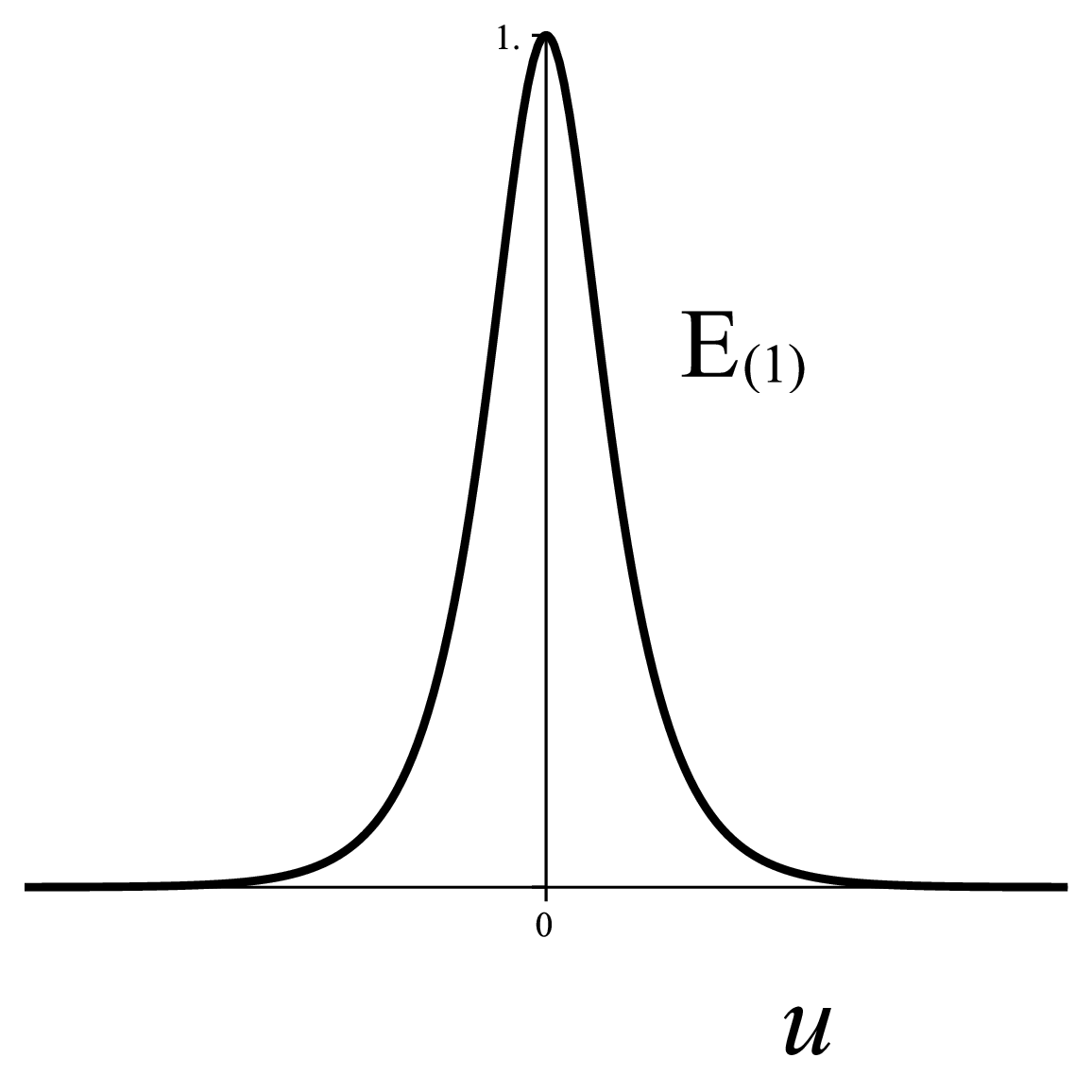}
\caption{The behavior of $E_{(1)}$  is shown for $\omega=1$ and $E_0=1$.}
\label{06022023b}
\end{figure}

\begin{figure}
\includegraphics[scale=0.3]{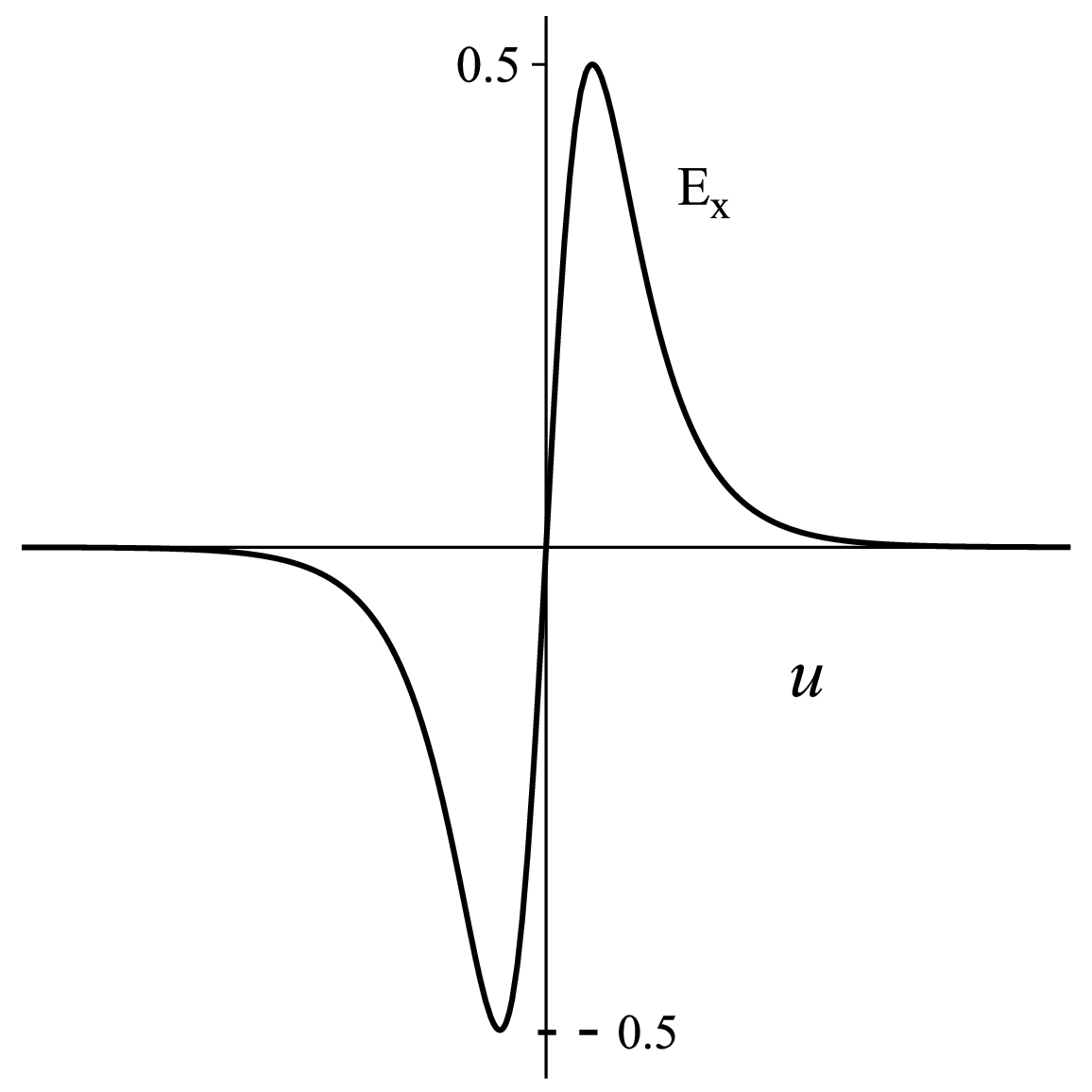}
\caption{The behavior of $E_x$  is shown for $\omega=1$ and $E_0=1$. The maximum and minimum occur at $\arctanh(\sqrt{2}/2)$ and $-\arctanh(\sqrt{2}/2)$, respectively. Their distance is $\Delta z=2(c/\omega)\arctanh(\sqrt{2}/2)$.}
\label{11122022a}
\end{figure}

Recalling that $E_{(1)}=E_0F_1$, we find that the maximum of $E_{(1)}$ occurs at $u=0$, as shown in Fig.\ref{06022023b}.
Now, recalling that $E_x=fE_{(1)}$, we find that $E_x=E_0\tanh(\theta)\sech(\theta)$. The maximum and minimum values of $E_x$ are $\pm E_0/2$ and they are at $\omega u_\pm=\pm \arctanh(\sqrt{2}/2)$. As expected, the source is located around $u=0$. In terms of $z$, we have $z_\pm=ct\pm(c/\omega)\arctanh(\sqrt{2}/2)$, where we are using the convention $u_\pm=t-z_\mp/c$.  The distance between the maximum and the minimum of $E_x$ is $2(c/\omega)\arctanh(\sqrt{2}/2)$. It is clear that the bigger $\omega$ is, the more localized the source is. The shape of $E_x$ is shown in Fig.~\ref{11122022a}.

\subsubsection{Energies}\label{25012023a}
Substitution of Eqs.~\eqref{21012023c} and \eqref{10012022a} into Eqs.~\eqref{20012023j} and \eqref{20012023l} yields $I_1=2/3$ and $I_2=4/3$, where we have used $z_>\to \infty$ and $z_<\to -\infty$. Assuming that the observer who measures $A$ is located at infinity, then we will have $f^2(u_0)=1$.  From Eq.~\eqref{20012023i}, we see that $\alpha_1=1/(6\pi)$ and $\alpha_2=1/(3\pi)$. Using these values and Eq.~\eqref{06022023a} in Eqs.~\eqref{20012023g}-\eqref{20012023h}, we find that  
\begin{align}
E_\mathrm{M}=-E_\mathrm{g}=c^3A\omega/(3\pi G).
\label{25012023b}
\end{align}
Note that the spacetime energy vanishes.

Now, if we assume that the energies are quantized, we find from Eq.~\eqref{20012023p} that $C_\mathrm{g}=C$. It is possible that this equality would imply that the number of gravitons equals that of photons. However, since we do not have a special reason to believe that the functional forms of $C_\mathrm{g}$ and $C$ in this particular spacetime  will yield this equality, we will not make this assumption here.  In any case, the quantum energies will be well defined only for areas with the values given by $A=3\pi C(N)l_p^2$ [see Eq.~\eqref{04022023a}].

We proceed now to the analysis of the frame.

\subsubsection{Analysis of the frame}
\begin{figure}
\includegraphics[scale=0.3]{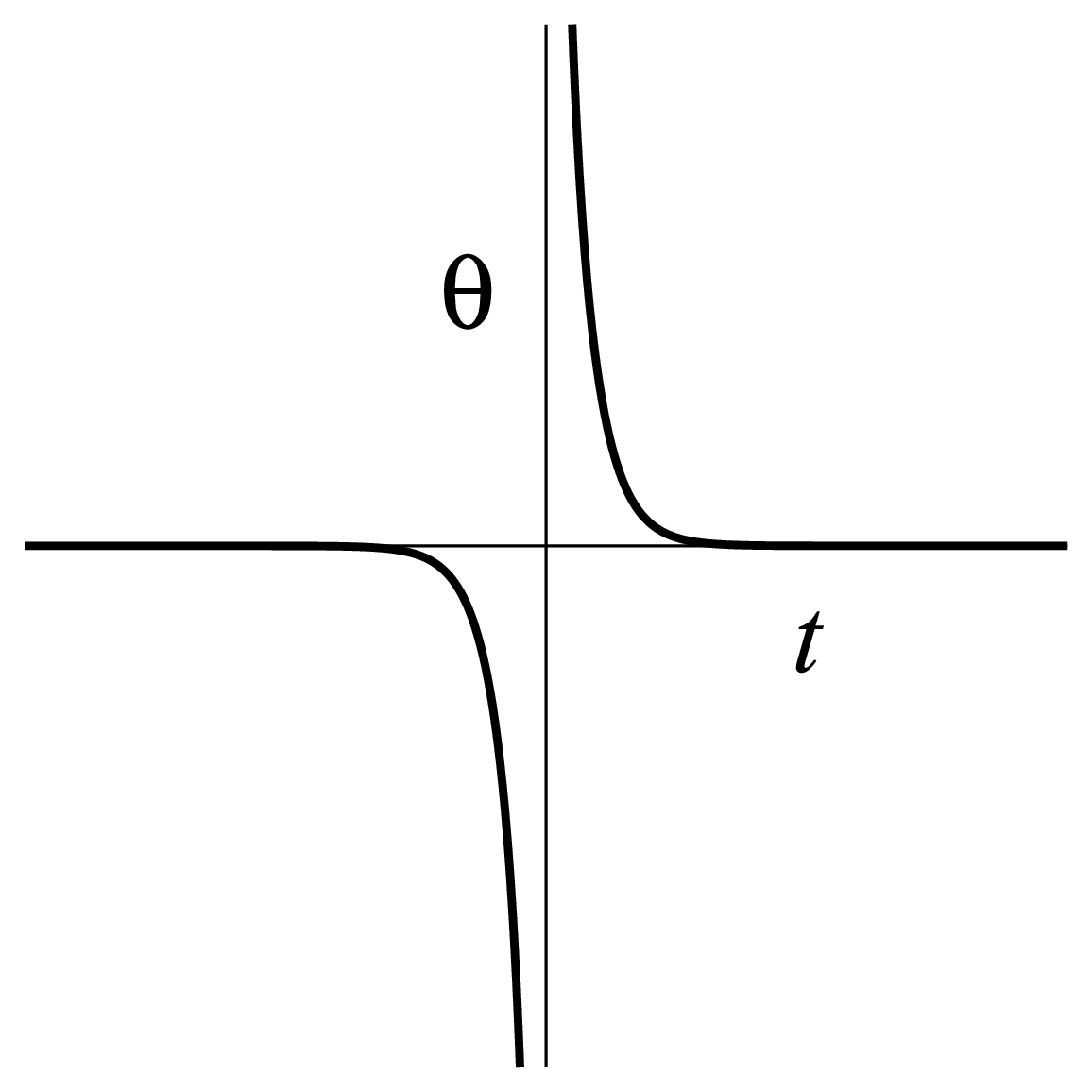}
\caption{The behavior of the expansion of the congruence, $\theta$, for the test particles in the plane $z=0$ as a function of $t$. The wave itself is at $z=ct$. The left side represent the expansion when the wave is approaching $z=0$, while the right side shows the behavior of $\theta$ after the wave has passed. In $t=0$, all the particles are at the same position, therefore producing a coordinate singularity.}
\label{29112022e}
\end{figure}

For this case, Eqs.~\eqref{29112022b} and \eqref{29112022c} give the expansion $\theta=2\omega/(\sinh(\omega u)\cosh(\omega u))$ and the shear tensor $\sigma_{(i)(j)}=(\theta/3)\left[(1/2)(\delta^1_i\delta^1_j+\delta^2_i\delta^2_j)-\delta^3_i\delta^3_j\right]$. The behavior of $\theta$ as a function of $t$ in the plane $z=0$ is sketched in Fig.~\ref{29112022e}. Since the wave is at $z=ct$, the negative values of $t$ in this figure represent the behavior of $\theta$ in the plane $z=0$ before the pulse has reached this plane,  while the region $t>0$ shows the behavior of the expansion after the pulse has passed. As it is clear in Fig.~\ref{29112022e}, as the wave gets closer and closer the volume of the fluid contracts ($\theta<0$); after the wave has passed, the opposite happens. In fact, there is a change of relative orientation between $\e{_{(1)}}$ and $\partial_x$, also between $\e{_{(2)}}$ and $\partial_y$ [see Eq.~\eqref{02012022a}], because $f$ changes its sign. (But $e$, the tetrad determinant, does not change sign.) It is clear that the coordinate system  becomes problematic at $u=0$. This occurs because the geodesic lines of the free particles intersect one another, $s_x=s_y=0$ for any $\Delta x$ and $\Delta y$. [This is a coordinate singularity. One can show this by using the coordinate transformation given by Eqs.~(4.4) and (4.9)-(4.10) in Ref.~\cite{griffiths1991colliding} to eliminate the singularity.]

Note that, although the frame given by Eq.~\eqref{02012022a} is a freely falling frame for the specific case considered here, it is not a local inertial frame of reference for any finite value of $u$. This is clear from the fact that $f^\prime\neq 0$ for $u$ finite [see Eq.~\eqref{10012022a}], i.e., the Christoffel symbols do not vanish, unless one takes the limit $u\to\pm\infty$.

\subsubsection{Energy densities}

We already know that the spacetime energy vanishes, in fact the $4$-momentum $P^a$ vanishes. Now we focus on the densities. As we will see, none of the densities vanish.

Substitution of Eqs.~\eqref{21012023c} and \eqref{10012022a}  into Eqs.~\eqref{26012023a} and \eqref{26012023b} yields
\begin{equation}
e{\cal T}^{\mu a}=\tilde{\rho}_0\tanh^2(\omega u)\sech^2(\omega u)\Delta^{\mu a}_{03},
\label{10012022d}
\end{equation} 
\begin{equation}
e\energy{^\mu^a}=-\frac{\tilde{\rho}_0}{2}\sech^4(\omega u)\Delta^{\mu a}_{03},
\label{10012022c}
\end{equation}
where we have used Eq.~\eqref{06022023a}. It is clear that $\energy{^\mu^a}+{\cal T}^{\mu a}\neq 0$. Thus, there is an unbalance between the gravitational and the electromagnetic energies ``locally''. 
 
From the definition of energy density given by Eq.~\eqref{04122022c} and the frame components given by Eq.~\eqref{02012022b}, we see that $\rho_\mathrm{g}=\energy{^0^{(0)}}$ and $\rho_\mathrm{M}={\cal T}^{0 (0)}$; of course, $\rho=\rho_\mathrm{g}+\rho_\mathrm{M}$. From Eqs.~\eqref{10012022d} and \eqref{10012022c}, we find that
\begin{align}
\tilde{\rho}=\tilde{\rho}_0\left[-\frac{1}{2}+\sinh^2(\omega u) \right]\sech^4(\omega u),
\label{10122022b}\\
\tilde{\rho}_\mathrm{g}=-\frac{\tilde{\rho}_0}{2}\sech^4(\omega u),
\label{10122022c}\\
\tilde{\rho}_\mathrm{M}=\tilde{\rho}_0\sinh^2(\omega u)\sech^4(\omega u).
\label{10122022d}
\end{align}
The pressures and the heat flow are obtained by substituting Eqs.~\eqref{10122022b}-\eqref{10122022d} into Eqs.~\eqref{05122022a}-\eqref{06122022a}. (Note that they are proportional to the energy densities.)

\begin{figure}
\includegraphics[scale=0.3]{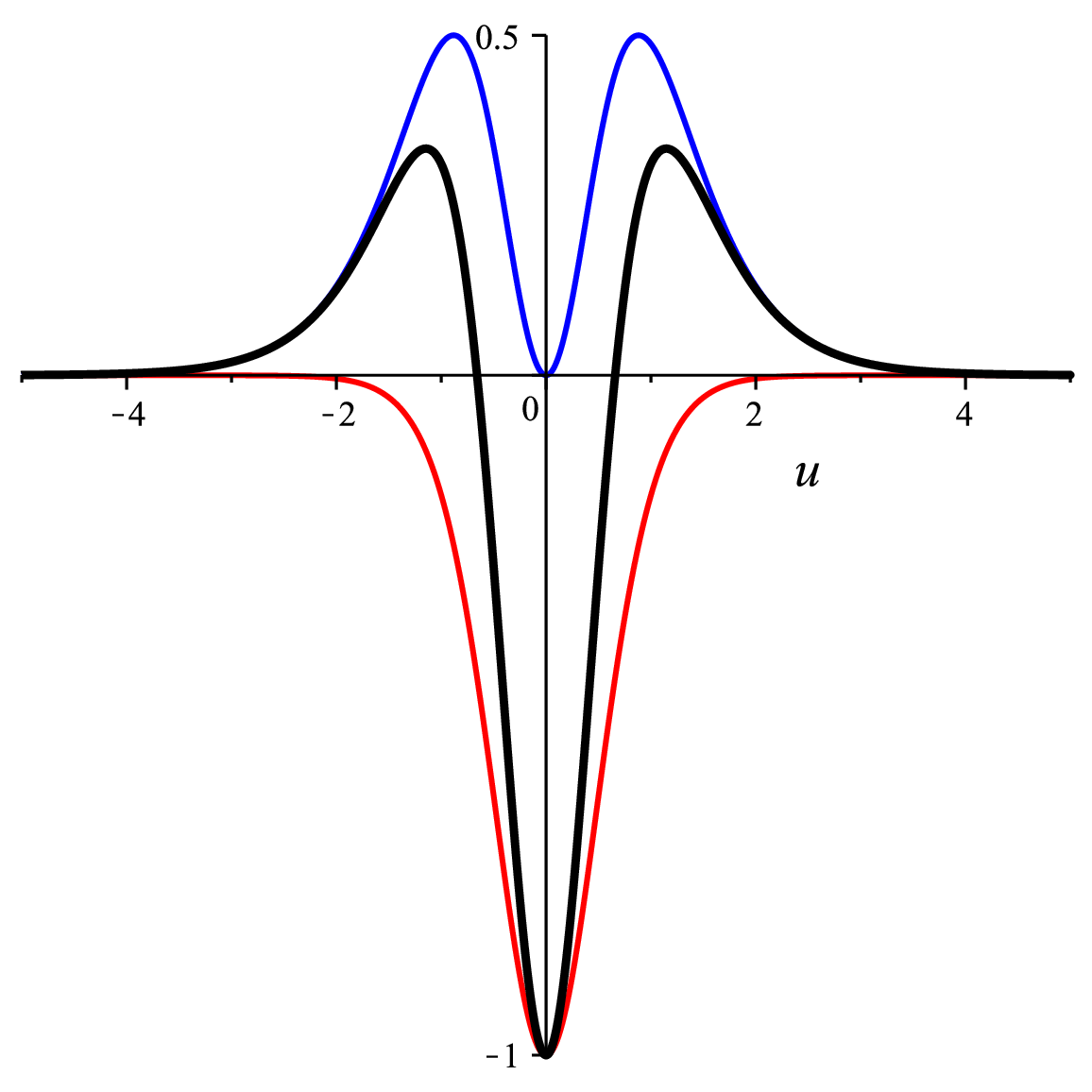}
\caption{The behavior of Eqs.~\eqref{10122022b}-\eqref{10122022d}. The curve with positive values (blue curve) corresponds to $\rho_M$, while the one with negative values (the red curve) is $\rho_g$; the spacetime energy density, $\rho$, is represented by the black curve. The spacetime energy density is positive in the region $|u|>\arcsinh(1/\sqrt{2})/\omega$. (In this plot, $\omega=1$ and $\tilde{\rho}_0=2$.)}
\label{10122022a}
\end{figure}

The behavior of the energy densities is sketched in Fig.~\ref{10122022a}. The maxima of $\tilde{\rho}_M$ occur at $\omega u_\pm=\pm\arcsinh(1)$ and their values are exactly $\tilde{\rho}_0/4$. On the other hand, the maxima of $\rho$ occur at $\omega u_\pm=\pm\arcsinh(\sqrt{2})$ and their values are precisely $\tilde{\rho}_0/6$. The gravitational energy density reaches its minimum value at $u=0$ and tends to zero as $u$ goes to $\pm \infty$.  

Notice that the spacetime energy density is positive only for $|u|>\arcsinh(1/\sqrt{2})/\omega\approx 0.7/\omega$; by taking $t=0$, this would given a distance of $|z|>0.7c/\omega\approx 0.1\lambda$ from the pulse, where $\lambda\equiv 2\pi c/\omega$. For a $\lambda$ comparable to the wavelength of visible light, let us say $\lambda=700\,$nm, this gives $|z|\gtrsim 10^{-7}$m.

\subsection{Periodic electromagnetic wave}\label{22012023b} 
Consider now the case in which  $E_{(1)}=E_0$, i.e., the electric field in the tetrad basis is constant. Equation \eqref{090102022a} becomes simply the harmonic oscillator equation  $f^{\prime\prime}=-\omega_0^2f$, and we take the solution as
\begin{align}
f=\cos(\omega_0 u), \quad (F_2=\cos \theta)
\label{22012023a}
\end{align}
where $\theta=\omega_0 u$; obviously, we have $\omega=\omega_0$ in this case. From Eq.~\eqref{20012023c} we see that the choice $E_{(1)}=E_0$ leads to $F_1=1$.

What is interesting about this choice and which justifies the name ``periodic electromagnetic wave'' is that the component $E_x$ is given by $E_x=E_0\cos(\omega_0 u)$. In other words, the electric field component in the coordinate basis of the free test particles oscillates with the same frequency as that of the metric components, and is also in phase with the metric. (Recall that $E_x=fE_{(1)}$.)

Some might say that those oscillations are meaningless because they are coordinate dependent. In order to see that this is not the case here, remember that the proper distance between two particles located along the $x$ axis is $s_x=f\Delta x=\cos(\omega_0 u)\Delta x$. (A similar expression holds for $s_y$) Furthermore, as we will see later on, the energy densities will oscillate too. Therefore, this oscillation is a fundamental property of the solution.

\subsubsection{Energies}
To calculate the energies, we use $z_>=z_<+n\lambda_0$, where $\lambda_0\equiv 2\pi c/\omega_0$ and $n=1,2,3,\ldots$. This means that we are going to calculate the energies inside a box with a length of $n\lambda_0$ along $z$ and a cross-sectional area $A$. 

Evaluating Eqs.~\eqref{20012023j}-\eqref{20012023l}, we obtain $I_1=I_2=n\pi$ and $\alpha_1=\alpha_2$. From this equality, Eqs.~\eqref{20012023g}-\eqref{20012023h}, and  $\omega_0=\omega$, we see that the spacetime energy vanishes. Therefore, the hypothesis that  the electromagnetic energy is quantized will lead to Eq.~\eqref{04022023a}.

Before we analyze the consequences of Eq.~\eqref{04022023a}, let us calculate the energies first. To measure the area $A$, we choose an observer that is located at $u_0=0$ at an instant $t$, which leads to $f(u_0)=1$. So, Eq.~\eqref{20012023i} yields $\alpha_1=\alpha_2=n/4$. Substitution into Eqs.~\eqref{20012023g}-\eqref{20012023h} yields the energies
\begin{align}
E_\mathrm{M}=-E_\mathrm{g}=\frac{c^3}{4G}nA\omega_0.
\label{22012023c}
\end{align}

Now, substituting the value of $\alpha_2$ into Eq.~\eqref{04022023a}, we find that 
\begin{align}
A=4C(N)\frac{l_p^2}{n}.
\label{22012023d}
\end{align}
We can also calculate the volume $V=\int d^3x e$ and find that $V=(1/2)n\lambda_0 A$. So, using Eq.~\eqref{22012023d} we arrive at
\begin{align}
V=2C(N)\lambda_0l_p^2.
\label{11022023a}
\end{align}
Note that the volume of the box makes no reference to spacetime parameters, such as $A$ and $n$.\footnote{Recall that the length of the box is $\Delta z=n\lambda$. So, $n$ is a free parameter that is basically measuring distances along $z$.}

We believe that we can interpret Eq.~\eqref{22012023d} in the following way. By using the set of classical test particles that make up the reference frame, the observer can measure any cross-sectional area. However, if the observer measures an area that cannot satisfy \eqref{22012023d}, then either the spacetime geometry is not given by Eq.~\eqref{22012023a}, in which case the observer's choice simply changed the spacetime geometry, or the energies are not well defined. Since from the classical point of view test particles do not change the geometry, the latter conclusion is more suitable here. So, we will stick to the latter. 

We are not literally saying that the area should be given exactly by Eq.~\eqref{22012023d}, in order to ensure that the energy is well defined. We are saying that this semi-classical approach is suggesting that the quantization of all non-gravitational fields require the discretization of geometry for consistency. The exact value should be given by a quantum theory of gravity, and not by the semi-classical approach considered here. [ We would not be surprised, however, if the right theory gave the relation \eqref{22012023d} in some approximation.]

Nevertheless, let us take Eqs.~\eqref{22012023d} and \eqref{11022023a} seriously and see what the implications are. For the sake of completeness, we will assume that $C(N)=N+1/2$, in analogy with a quantum harmonic oscillator. In this case, the area and the volume are related to the number of photons by 
\begin{align}
A=4\left(N+\frac{1}{2}\right)\frac{l_p^2}{n},
\label{23012023a}
\end{align}
\begin{align}
V=2\left(N+\frac{1}{2}\right)l_p^2\lambda_0.
\label{11022023b}
\end{align}

Following the interpretation above, we could say that the smallest area and the smallest volume for a well-defined vacuum energy in the spacetime \eqref{22012023a} are $A=2l_p^2/n$ and  $V=\lambda_0 l_p^2$, respectively. Note that the longer the box, the smaller $A$ becomes. Curiously enough, one finds that $V$ equals the Planck volume if $\lambda_0=l_p$.\footnote{From Eq.~\eqref{20012023e}  one can calculate the amplitude of the electric field associated with $\lambda_0=l_p$. This gives $E_0\sim 10^{62}$V/m, which is a huge value.}

Of course, the observer could use the test particles of the frame to measure arbitrarily smaller areas without changing the spacetime geometry. The question, however, is whether this would be possible if the field describing these particles were quantized. It seems that a state with well-defined energy would be related to a state with well-defined area and volume, and the area and the volume would be discrete.

From Eqs.~\eqref{23012023a} and \eqref{11022023b}, we can see that $A_\mathrm{N+1}-A_\mathrm{N}=(4/n)l_p^2$ and $V_\mathrm{N+1}-V_\mathrm{N}=2\lambda_0l_p^2$, which suggest that each additional photon requires an increasing of $(4/n)l_p^2$ in the cross-sectional area and an increasing of $2\lambda_0l_p^2$ in the volume. We also find that the density of photons is $N/V\approx 1/(2\lambda_0 l_p^2)\approx 2\times 10^{69}\textrm{[photons/m$^2$]}/\lambda_0$, for $V\gg\lambda_0l_p^2$, which is a high density for any ``reasonable'' wavelength.

\subsubsection{Analysis of the frame}

The expansion tensor of the congruence $x^j$=constant (timelike geodesics) is $\theta_{(i)(j)}=-\omega_0\tan(\omega_0 u)(\delta^1_i\delta^1_j+\delta^2_i\delta^2_j)$. So, the expansion is $\theta=-2\omega_0\tan(\omega_0 u)$, and the shear tensor (traceless part) is $\sigma_{(i)(j)}=(\omega_0/3)\tan(\omega_0 u)\left(-\delta^1_i\delta^1_j-\delta^2_i\delta^2_j+2\delta^3_i\delta^3_j \right)$. As expected, the volume keeps increasing and decreasing periodically, and there is a coordinate singularity\footnote{Again, we can prove that they are coordinate singularities by removing them with the help of the coordinate transformation given by Eqs.~(4.4) and (4.9)-(4.10) in Ref.~\cite{griffiths1991colliding}.} every time $\omega_0 u$ equals $m\pi/2$, for $m$ odd. (Note that $f$ vanishes in these cases.)  

Unlike the asymptotically flat frame of Sec.\ref{01012023b}, here we have a frame that is a local inertial frame at the events where $u=0$: from Eq.~\eqref{22012023a}, we see that the metric components become that of Minkowski and its first derivative vanishes when $u=0$. Since $\e{_a}$ is adapted to the coordinates $(ct,x,y,z)$, this means that $\e{_a}$ becomes $\e{_a}=(\partial_{x^0},\partial_x,\partial_y,\partial_z)$ and the connection coefficients vanish when $u=0$ [see, e.g., Eqs.~\eqref{02012022a} and \eqref{28112022c}]. As we will see, the gravitational stress-tensor do vanish at these events\footnote{It is interesting to see that in this particular case the gravitational energy naturally vanishes at these events: there is no need to add any non-gravitational interaction to make $\rho_\mathrm{g}$ vanish.}.

Finally, one can easily check that the limit \eqref{28012023a} is satisfied. So, in the absence of gravity, the frame becomes a global inertial frame of reference.

\subsubsection{Energy densities}
Recalling that $F_1=1$ and using \eqref{22012023a} in Eqs.~\eqref{26012023a}-\eqref{26012023b}, we find that
\begin{equation}
e{\cal T}^{\mu a}=\tilde{\rho}_0\cos^2(\omega_0 u) \Delta^{\mu a}_{03},
\label{11012022d}
\end{equation}
\begin{equation}
e\energy{^\mu^a}=-\tilde{\rho}_0\sin^2(\omega_0 u) \Delta^{\mu a}_{03},
\label{11012022e}
\end{equation}
where we have used the fact that $\omega=\omega_0$, and also Eqs.~\eqref{20012023e} and \eqref{26012023c}. The energy densities are

\begin{align}
\tilde{\rho}_\mathrm{M}=\tilde{\rho}_0\cos^2(\omega_0 u),
\label{28122022c}
\\
\tilde{\rho}_\mathrm{g}=-\tilde{\rho}_0\sin^2(\omega_0 u),
\label{28122022b}
\\
\tilde{\rho}=\tilde{\rho}_0\cos(2\omega_0 u).
\label{28122022a}
\end{align}
As expected, $\tilde{\rho}_\mathrm{g}$ vanishes when $u=0$.

\begin{figure}
\includegraphics[scale=0.3]{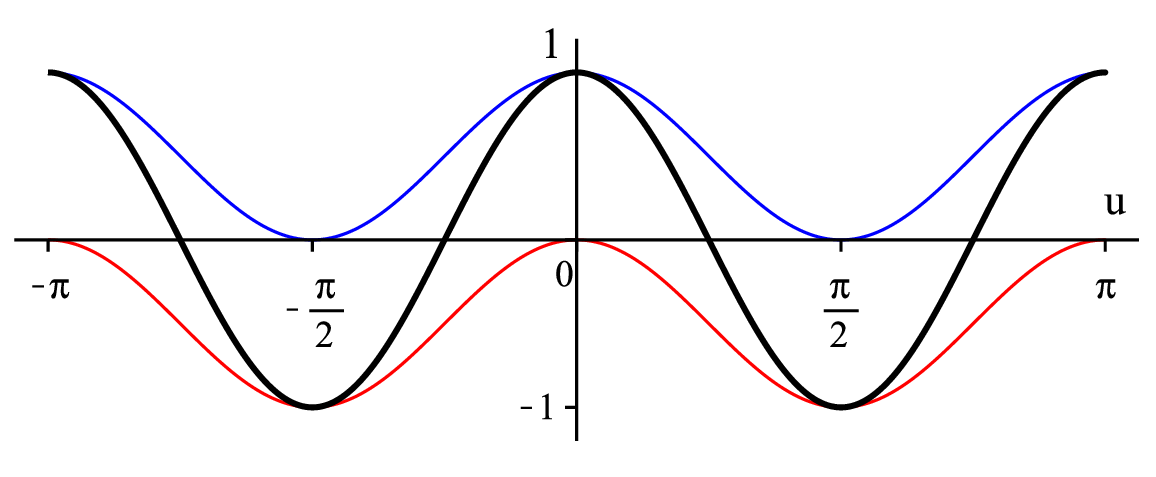}
\caption{The behavior of Eqs.~\eqref{28122022c}-\eqref{28122022a}. The curve with positive values (blue curve) corresponds to $\rho_\mathrm{M}$, while the one with negative values (the red curve) is $\rho_\mathrm{g}$; the spacetime energy density, $\rho$, is represented by the black curve. The spacetime energy density is positive in the regions $(2m-1)\pi/(4\omega_0)< u < (2m+1)\pi/(4\omega_0)$, for $m$ even. (We are taking $\tilde{\rho}_0=1$ and $\omega_0=1$. )} 
\label{27122022a}
\end{figure}

The behavior of $\tilde{\rho}$, $\tilde{\rho}_\mathrm{g}$, and $\tilde{\rho}_\mathrm{M}$ are sketched in Fig.~\ref{27122022a}. All these densities reach their maxima at $u=0,\pm \pi,\pm 2\pi,\ldots$. In turn, the positive values of $\tilde{\rho}$ occur in the regions $(2m-1)\pi/(4\omega_0)< u < (2m+1)\pi/(4\omega_0)$, for $m$ even.

The periodic solution we obtained here is a very special case of more general solutions. For instance, by assuming that $F_1=(\omega/\omega_0)\sqrt{\lambda-2q\cos(2\theta)}$, where $\lambda$ and $q$ are parameters such that $\lambda-2q\cos(2\theta)\geq 0$ for all $\theta$, Eq.~\eqref{21012023a} becomes the well-known angular Mathieu equation [See, e.g., Eq.~(13.181) of Ref.~\cite{Arfken}.], which admits periodic solutions. (The case treated here can be obtained from $\lambda=\omega_0^2/\omega^2$ and $q=0$.)

\section{Concluding remarks}\label{30012023b}
We have seen that the TEGR gives a consistent description of the stress-energy tensors for the solutions of the Einstein-Maxwell equations known as $pp$-waves, at least for the case of plus polarization. In doing so, we have decomposed the stress tensors and shown that the energy densities, the pressures, and the heat flows are all consistent with the electromagnetic wave. For example, we showed that the heat flow exhibits a energy flux along the $z$ direction, as expected.

This consistency was achieved by using a tetrad field that is adapted to test particles in free fall. We believe that the frames we used here are free from artificial properties and reflect only the effects of gravity, with the possible exception of the asymptotically flat case.  It would be interesting though to study other types of frame that are also free from artificial properties but have nongravitational interactions. This study could give us a better understanding of how those interactions change the meaning of $\energy{^\mu^a}$.

The results obtained in this paper suggest that ``area'' plays a fundamental role in gravity. We showed that the spacetime energy of the $pp$-waves, with plus polarization, and described in a freely falling frame, can be written in terms of the proper areas measured by two different observers located at opposite faces of a box. When the rate of change of these areas are equal, the spacetime energy vanishes. We have also been able to write the gravitation and electromagnet $4$-momenta in terms of the proper area measured by an observer; in the ``asymptotically flat'' case, the observer was far away from the source, while in the periodic case it was at the center of the box. In all cases, area was playing an important role.

Another interesting role played by these areas was the relations that they had to satisfy when we considered that the electromagnetic energy is quantized. We found that these areas had also to be quantized. Since the solutions of Einstein-Maxwell equations are ``blind'' to test particles, at least from the perspective of invariant properties such as the spacetime curvature, one can always add test particles to continuously measure areas without changing the spacetime geometry. However it seems that, for consistency with the quantization of the electromagnetic energy, the boxes cannot have a well defined number of photons, if the measured areas differ from those of the quantized version.

Now, in quantum mechanics, the concept of ``test particles'' is problematic \cite{Okon2011}. We may not be able to build a frame of reference with particles whose energies are as small as desired, in order to neglect the stress-energy tensor of the constituents of the frame (as is usually done in the classical case). Furthermore, if all matter fields are quantized, including the particles of the frame, one may find that the measured area must always satisfy a relation like that of Eq.~\eqref{23012023a}. In this case, it would be natural to think that the quantization of area is indeed a fundamental property of the spacetime. 

However, we do not know  whether this type of relation is just a particular feature of the solutions we obtained. In the worked examples, we considered only cases where $E_0$ is uniquely determined by $\omega$. It may be the case that other solutions do not relate the area to a quantized quantity. 

A curious feature of the toy model considered here is that, although we have not localized a photon at a specific spacetime point, we were able to localize it in a specific volume for the periodic case. For this solution, we saw that $N$ photons ``occupy'' the volume $V=2(N+1/2)l_p^2\lambda_0$. This is a kind of localization because one photon of wavelength $\lambda_0$ would be inside the volume $V=3l_p^2\lambda_0$ that is centered at $z_0$. 

There is another important point to be made here. Locality is one of those words that are frequently used, sometimes even with different meanings, most people think they know exactly what it is but, in fact, it seems that no one really knows exactly what it means. Although our intuition of ``locality'' works very well in most cases, when it comes to the principle of equivalence it does not seem to be satisfactory. (See, e.g., Refs.~\cite{doi:10.1119/1.10744,OhanianBook,NORTON1985203,doi:10.1119/1.4895342}.) In this paper, we have expressed the gravitational energy-momentum tensor in terms of quantities that, in principle, can be measured by a freely falling observer. We use this result to discuss its implications on the problem of localizing the gravitational energy, and argued that it gives support to the idea that one can build a consistent stress tensor for the gravitational field regardless of what the word ``locality'' really means.

As is well known, the matter stress-energy tensor is supposed to satisfy the so called energy conditions, both in GR and in extended theories of gravity \cite{PhysRevD.91.124019}. It is not clear, though, whether the gravitational stress-energy tensor (and the spacetime one) should satisfy similar conditions. Any condition imposed on them will limit either the number of possible tetrad fields or solutions of Einstein's equations. For example, if we demand that the gravitational energy density be positive, then either the tetrads used here are not suitable or the $pp$-wave solutions we considered cannot exist in the real world.

Concerning the experimental test of quantum gravity, it has been shown in Ref.~\cite{PhysRevLett.93.191301} that preferred-frame effects in quantum gravity can give rise to detectable Lorentz violation. Based on this result, one may claim that the model adopted here suffers from this problem. However, we do not think that this is the case. First, TEGR field equations do not depend on the tetrad field; we have used the freely falling frame for convenience, not because the theory is frame dependent. Second, the authors in Ref.~\cite{PhysRevLett.93.191301}  approach the problem of Lorentz violation by adding a new term, $\Pi(E,\bm{p})$, to the relativistic dispersion relation, i.e., they write $P^\mu P_\mu=m^2+\Pi(E,\bm{p})$. In the present paper, even after the ``quantum hypothesis'', all $4$-momenta continue to satisfy the relation  $P^aP_a=0$, which is invariant under global Lorentz transformations.  Furthermore, as far as we know, the accelerated frames also yield the relativistic dispersion relation. (See, e.g., section 4 of Ref. [36].) 

The results obtained in this paper are limited to waves with constant linear polarization.  We believe, however, that the extension to the general case of $pp$-waves will not change the mains results.


\section*{Acknowledgments}
Jo\~ao A. C. Duarte acknowledges PIBIC/UFPB for financial support.

%


\end{document}